\documentclass[preprintnumbers,superscriptaddress,amsmath,amssymb,twocolumn,aps,prc,showpacs,nofootinbib]{revtex4-2}
\usepackage{graphicx}
\usepackage{dcolumn}
\usepackage{bm}

\usepackage[chatter]{rotating}
\def\bea{\begin{eqnarray}}
\def\eea{\end{eqnarray}}

\begin{document}

\preprint{Version 8}

\title{Two-particle correlations on transverse rapidity in Au+Au collisions at $\sqrt{s_{\rm NN}}=200$ GeV at STAR}

\affiliation{Abilene Christian University, Abilene, Texas   79699}
\affiliation{AGH University of Science and Technology, FPACS, Cracow 30-059, Poland}
\affiliation{Alikhanov Institute for Theoretical and Experimental Physics NRC "Kurchatov Institute", Moscow 117218}
\affiliation{Argonne National Laboratory, Argonne, Illinois 60439}
\affiliation{American University of Cairo, New Cairo 11835, New Cairo, Egypt}
\affiliation{Ball State University, United States}
\affiliation{Brookhaven National Laboratory, Upton, New York 11973}
\affiliation{University of Calabria \& INFN-Cosenza, Italy}
\affiliation{University of California, Berkeley, California 94720}
\affiliation{University of California, Davis, California 95616}
\affiliation{University of California, Los Angeles, California 90095}
\affiliation{University of California, Riverside, California 92521}
\affiliation{Central China Normal University, Wuhan, Hubei 430079 }
\affiliation{University of Illinois at Chicago, Chicago, Illinois 60607}
\affiliation{Creighton University, Omaha, Nebraska 68178}
\affiliation{Czech Technical University in Prague, FNSPE, Prague 115 19, Czech Republic}
\affiliation{Technische Universit\"at Darmstadt, Darmstadt 64289, Germany}
\affiliation{ELTE E\"otv\"os Lor\'and University, Budapest, Hungary H-1117}
\affiliation{Frankfurt Institute for Advanced Studies FIAS, Frankfurt 60438, Germany}
\affiliation{Fudan University, Shanghai, 200433 }
\affiliation{University of Heidelberg, Heidelberg 69120, Germany }
\affiliation{University of Houston, Houston, Texas 77204}
\affiliation{Huzhou University, Huzhou, Zhejiang  313000}
\affiliation{Indian Institute of Science Education and Research (IISER), Berhampur 760010 , India}
\affiliation{Indian Institute of Science Education and Research (IISER) Tirupati, Tirupati 517507, India}
\affiliation{Indian Institute Technology, Patna, Bihar 801106, India}
\affiliation{Indiana University, Bloomington, Indiana 47408}
\affiliation{Institute of Modern Physics, Chinese Academy of Sciences, Lanzhou, Gansu 730000 }
\affiliation{University of Jammu, Jammu 180001, India}
\affiliation{Joint Institute for Nuclear Research, Dubna 141 980}
\affiliation{Kent State University, Kent, Ohio 44242}
\affiliation{University of Kentucky, Lexington, Kentucky 40506-0055}
\affiliation{Lawrence Berkeley National Laboratory, Berkeley, California 94720}
\affiliation{Lehigh University, Bethlehem, Pennsylvania 18015}
\affiliation{Max-Planck-Institut f\"ur Physik, Munich 80805, Germany}
\affiliation{Michigan State University, East Lansing, Michigan 48824}
\affiliation{National Research Nuclear University MEPhI, Moscow 115409}
\affiliation{National Institute of Science Education and Research, HBNI, Jatni 752050, India}
\affiliation{National Cheng Kung University, Tainan 70101 }
\affiliation{Nuclear Physics Institute of the CAS, Rez 250 68, Czech Republic}
\affiliation{Ohio State University, Columbus, Ohio 43210}
\affiliation{Institute of Nuclear Physics PAN, Cracow 31-342, Poland}
\affiliation{Panjab University, Chandigarh 160014, India}
\affiliation{NRC "Kurchatov Institute", Institute of High Energy Physics, Protvino 142281}
\affiliation{Purdue University, West Lafayette, Indiana 47907}
\affiliation{Rice University, Houston, Texas 77251}
\affiliation{Rutgers University, Piscataway, New Jersey 08854}
\affiliation{Universidade de S\~ao Paulo, S\~ao Paulo, Brazil 05314-970}
\affiliation{University of Science and Technology of China, Hefei, Anhui 230026}
\affiliation{South China Normal University, Guangzhou, Guangdong 510631}
\affiliation{Shandong University, Qingdao, Shandong 266237}
\affiliation{Shanghai Institute of Applied Physics, Chinese Academy of Sciences, Shanghai 201800}
\affiliation{Southern Connecticut State University, New Haven, Connecticut 06515}
\affiliation{State University of New York, Stony Brook, New York 11794}
\affiliation{Instituto de Alta Investigaci\'on, Universidad de Tarapac\'a, Arica 1000000, Chile}
\affiliation{Temple University, Philadelphia, Pennsylvania 19122}
\affiliation{Texas A\&M University, College Station, Texas 77843}
\affiliation{University of Texas, Austin, Texas 78712}
\affiliation{Tsinghua University, Beijing 100084}
\affiliation{University of Tsukuba, Tsukuba, Ibaraki 305-8571, Japan}
\affiliation{United States Naval Academy, Annapolis, Maryland 21402}
\affiliation{Valparaiso University, Valparaiso, Indiana 46383}
\affiliation{Variable Energy Cyclotron Centre, Kolkata 700064, India}
\affiliation{Warsaw University of Technology, Warsaw 00-661, Poland}
\affiliation{Wayne State University, Detroit, Michigan 48201}
\affiliation{Yale University, New Haven, Connecticut 06520}

\author{M.~S.~Abdallah}\affiliation{American University of Cairo, New Cairo 11835, New Cairo, Egypt}
\author{B.~E.~Aboona}\affiliation{Texas A\&M University, College Station, Texas 77843}
\author{J.~Adam}\affiliation{Brookhaven National Laboratory, Upton, New York 11973}
\author{L.~Adamczyk}\affiliation{AGH University of Science and Technology, FPACS, Cracow 30-059, Poland}
\author{J.~R.~Adams}\affiliation{Ohio State University, Columbus, Ohio 43210}
\author{J.~K.~Adkins}\affiliation{University of Kentucky, Lexington, Kentucky 40506-0055}
\author{G.~Agakishiev}\affiliation{Joint Institute for Nuclear Research, Dubna 141 980}
\author{I.~Aggarwal}\affiliation{Panjab University, Chandigarh 160014, India}
\author{M.~M.~Aggarwal}\affiliation{Panjab University, Chandigarh 160014, India}
\author{Z.~Ahammed}\affiliation{Variable Energy Cyclotron Centre, Kolkata 700064, India}
% ...

\collaboration{STAR Collaboration}\noaffiliation

\date{\today}
\begin{abstract}

	Two-particle correlation measurements projected onto two-dimensional, transverse rapidity coordinates ($y_{T1},y_{T2}$), allow access to dynamical properties of the QCD medium produced in relativistic heavy-ion collisions that angular correlation measurements are not sensitive to. We report non-identified charged-particle correlations for Au + Au minimum-bias collisions at $\sqrt{s_{\rm NN}}$ = 200~GeV taken by the STAR experiment at the Relativistic Heavy-Ion Collider (RHIC). Correlations are presented as 2D functions of transverse rapidity for like-sign, unlike-sign and all charged-particle pairs, as well as for particle pairs whose relative azimuthal angles lie on the near-side, the away-side, or at all relative azimuth. The correlations are constructed using charged particles with transverse momentum $p_T \geq 0.15$~GeV/$c$, pseudorapidity from $-$1 to 1, and azimuthal angles from $-\pi$ to $\pi$. The significant correlation structures that are observed evolve smoothly with collision centrality. The major correlation features include a saddle shape plus a broad peak with maximum near $y_T \approx 3$, corresponding to $p_T \approx$ 1.5~GeV/$c$. The broad peak is observed in both like- and unlike-sign charge combinations and in near- and away-side relative azimuthal angles. The all-charge, all-azimuth correlation measurements are compared with the theoretical predictions of {\sc hijing} and {\sc epos}. The results indicate that the correlations for  peripheral to mid-central collisions can be approximately described as a superposition of nucleon + nucleon collisions with minimal effects from the QCD medium. Strong medium effects are indicated in mid- to most-central collisions.
\end{abstract}

\pacs{25.75.q,25.75.Gz}
\maketitle


\section{Introduction}
\label{SecI}

Two-particle correlation measurements in high-energy heavy-ion collisions provide access to partonic and hadronic dynamics occurring throughout the spatial and temporal evolution of the produced hot and dense matter. The dynamical processes include soft and hard interactions as predicted by Quantum Chromodynamics (QCD), hadronization via fragmentation~\cite{LUND,PYTHIA,TomFrag} and/or recombination~\cite{RudyHwa}, partonic and hadronic collective flow~\cite{flow}, resonance decays, quantum interference effects~\cite{HBT,Levin}, and others~\cite{Tomreview}.

Two-particle correlations in momentum space contain, in general, six independent coordinates. However, for identical, unpolarized colliding ions ({\em e.g.,} p+p, Au+Au, Pb+Pb) and for particle production near mid-rapidity, two-particle correlations can be accurately represented as functions of four variables $p_{T1}$, $p_{T2}$, relative pseudorapidity\footnote{Pseudorapidity is defined as $\eta = -\ln[\tan(\theta/2)]$, where $\theta$ is the polar scattering angle relative to the beam direction.} $\Delta\eta = \eta_1 - \eta_2$, and relative azimuthal angle $\Delta\phi = \phi_1 - \phi_2$ as in Refs.~\cite{AyaCD,LHCeta1eta2,Tompp}.
Correlation measurements on ($\Delta\eta,\Delta\phi$) angular space within a grid of bins on transverse momentum space ($p_{T1},p_{T2}$)~\cite{trigassoc,LizHQ,LizThesis,PrabhatThesis,Kettlerptdependent,KoljaPaper,Joern} represent all of the statistically accessible information available from the non-identified, two-particle distribution.

Two-particle correlation measurements on $\Delta\phi$ and/or $\Delta\eta$ are ubiquitous in the heavy-ion literature. However, much less attention has been given to correlations on transverse momentum dependent coordinates. The latter type of measurement was reported by the NA49 Collaboration~\cite{ReidNA49mtmt,NA49ptpt}, the CERES Collaboration~\cite{CERESptpt}, and the STAR Collaboration~\cite{Ayamtmt} (see also Refs.~\cite{ReidSTARmtmt,Tompp,Porter}). In this paper we present two-particle, 2D pair-number correlation distributions on transverse rapidity $(y_{T1},y_{T2})$ for minimum-trigger-biased Au+Au collisions at $\sqrt{s_{\rm NN}}$ = 200~GeV for various combinations of charge-sign, $\Delta\phi$ ranges, and covering cross-section fractions from 0\% to 93\% in eleven centrality bins. 
Transverse rapidity in this application is defined by
\bea
y_T = \ln[(p_T+m_T)/m_0],
\label{Eq0}
\eea
where $m_T = \sqrt{p_T^2 + m_0^2}$ is the transverse mass for particle mass $m_0$, assumed equal to the pion mass throughout this paper.\footnote{With this definition, $y_T \approx \ln{p_T} + \ln{(2/m_0)}$ within the $p_T$ range studied here, and equals $(1/2)\ln{[(E+p_T)/(E-p_T)]}$ when $\eta$ = 0.} Pions account for approximately 80\% of the charged particle multiplicity in this collision system~\cite{PRC79}.

The present analysis uses a correlation measure quantity~\cite{MCBias} that was derived from a minimum-statistically-biased mean-$p_T$ fluctuation quantity~\cite{meanptpaper}. The correlation distributions are projected onto the transverse rapidity defined in Eq.~(\ref{Eq0}) to facilitate studies of jet fragment contributions to these correlations~\cite{phenom}. The choice of transverse rapidity, with fixed pion-mass was, in part, based on the analysis in Ref.~\cite{TomFrag}. This analysis showed that non-identified particle, jet-fragment distributions produced in high-energy collisions, when plotted as functions of $y_T$ with fixed pion mass, displayed approximate shape invariance over a wide energy-range. The use of coordinate $y_T$ also enables better visual access to the correlation structures at both lower and intermediate momentum. Many correlation distributions are contained in this analysis corresponding to various charge-sign, $\Delta\phi$ range, and centrality combinations, thereby increasing the wealth of such correlation data in the heavy-ion literature. Representative examples are shown here.

Studies of correlation distributions on transverse momentum coordinates independently access different manifestations of heavy-ion collision dynamics beyond that observed in angular correlations~\cite{phenom}. For example, in the hydrodynamic picture, event-wise fluctuations in an equilibrated, global temperature~\cite{Baier,Kurkela} would not be evident in angular correlations, but would produce a distinctive ``saddle-shape'' correlation distribution on transverse momentum coordinates~\cite{NA49ptpt,Ayamtmt,phenom}. In fragmentation-based models with jets, {\em e.g.,} {\sc hijing}~\cite{HIJING}, where event-wise fluctuations occur in the angular positions and energies of the jets, analysis of angular correlations can determine the average total number of jet-related pairs of particles per event. On the other hand, analysis of correlations on transverse momentum dependent coordinates can determine the variance in the fluctuating number of jet-related pairs due to both the varying number and energies of the jets produced in each event. The latter represents independent information about jet production and fragmentation.

Interpretation of angular correlations is relatively straightforward because the principal structural features display simple geometrical shapes that can be described using the first few terms in an azimuthal cosine-series plus Gaussians for the peaks. These geometrical structures can be readily modelled with hydrodynamic, fragmentation, jet models, Bose-Einstein or Hanbury-Brown and Twiss (HBT)~\cite{HBT} correlations, and other models~\cite{axialCI}. On the other hand, the correlation distributions on transverse rapidity reported here cannot be readily decomposed into simple geometrical structures, nor do the observed features offer straightforward physical interpretation. Event-wise dynamical fluctuations that alter the shape of the underlying single-particle parent $y_T$-distribution, {\em e.g.,} fluctuations in temperature, transverse flow, jet production, and color-string energies, give rise to non-zero correlation distributions on transverse rapidity. The resulting structures display maxima and minima that form a generic saddle-shape. Different dynamical contributions, for example as studied in Ref.~\cite{phenom}, produce differing saddle-shapes.

To provide theoretical context, the correlation predictions of {\sc hijing}~\cite{HIJING}, in which a superposition of nucleon + nucleon (NN) collisions is assumed, and predictions of the (3+1)-dimensional hydrodynamic code {\sc epos}~\cite{EPOS} are compared with the data. The major features of the observed and predicted correlation structures are compared in detail. Theoretical analysis of the correlations presented here in combination with corresponding angular correlations~\cite{axialCI} may help distinguish between models and guide their development. This may lead to a better understanding of heavy-ion collision dynamics.

This paper is organized as follows. The correlation analysis method is described in Sec.~\ref{SecII}. Details of the experimental data and event processing are expatiated in Sec.~\ref{SecIII}. Examples of the measured correlations are shown and discussed in Sec.~\ref{SecIV} and the associated systematic uncertainties are discussed in Sec.~\ref{SecV}. The theoretical model comparisons and the physical implications are discussed in Secs.~\ref{SecVI} and \ref{SecVII}. A summary and conclusion are given in Sec.~\ref{SecVIII}. Further details of the analysis are provided in the appendices. 

\section{Analysis method}
\label{SecII}  

The two-particle correlations in this paper are derived from the normalized (within the range [$-$1,1]) covariance~\cite{Pearson} given by
\bea
\frac{ \langle \left( n_{k1} - \langle n_{k1} \rangle \right) \left( n_{l2} - \langle n_{l2} \rangle \right) \rangle }
{ \sqrt{ \sigma^2_{k1} \sigma^2_{l2} } } \nonumber \\
 &  & 
\hspace{-1.0in} \approx \frac{ \langle n_{k1} n_{l2} \rangle - \langle n_{k1} \rangle \langle n_{l2} \rangle } { \sqrt{ \langle n_{k1} \rangle \langle n_{l2} \rangle }}
\label{Eq1}
\eea
where $n_{k1}$ and $n_{l2}$ are the number of particles in single-particle bins $k$ and $l$ on transverse rapidity, $n_{k1} n_{l2}$ is the number of particle pairs in bin $(k,l)$, subscripts 1 and 2 are particle labels, $\sigma^2$ is the variance of the event-wise distribution of particle number in a single-particle bin. Brackets $(\langle \cal{O} \rangle )$ indicate averages over all collision events in the multiplicity or centrality bin. In the last line of Eq.~(\ref{Eq1}) the Poisson limit was assumed where $\sigma^2_k = \langle n_k \rangle$. This normalized covariance is bounded between $-1$ and $+1$ regardless of the event multiplicity, where the amplitude indicates whether the particle pairs in bins $(k,l)$ are fully correlated (+1), anti-correlated ($-1$), or somewhere in-between. 

The above ratio may be rewritten as
\bea
\sqrt{ \langle n_{k1} \rangle \langle n_{l2} \rangle }
\frac{ \langle n_{k1} n_{l2} \rangle - \langle n_{k1} \rangle \langle n_{l2} \rangle } { \langle n_{k1} \rangle \langle n_{l2} \rangle } \nonumber \\
 &  & \hspace{-1.0in} \equiv {\cal P}_{kl}
\frac{\rho_{{\rm se},kl} - \rho_{{\rm me},kl}} {\rho_{{\rm me},kl}}
\label{Eq2a}
\eea
where new symbols on the right-hand side (RHS) of Eq.~(\ref{Eq2a}) represent the corresponding event-average quantities on the left-hand side of this equation. ${\cal P}_{kl}$ is a prefactor, discussed at the end of this section and in Appendix~\ref{Appendix-A}. Quantities $\rho_{{\rm se},kl}$ and $\rho_{{\rm me},kl}$ are the average number of particle pairs in bin $(k,l)$ where pairs are from the same-event (se) and mixed-events (me), respectively. Particle labels 1 and 2 are omitted for brevity. The steps going from Eq.~(\ref{Eq1}) to Eq.~(\ref{Eq2a}) emphasize the essential nature of the prefactor that ensures normalization and insensitivity to system size. 

The form of the normalized covariance given in Eq.~(\ref{Eq2a}) is necessary for data analysis where particle reconstruction efficiency and acceptance effects cancel to first-order in the ratio term on the RHS when the mixed-event and same-event pair quantities $\rho_{{\rm me},kl}$ and $\rho_{{\rm se},kl}$ are constructed from similar events (see Sec.~\ref{SecIII}). Further corrections are required for this ratio as discussed in Sec.~\ref{SecIII}. Efficiency and acceptance corrections are also required for the prefactor as discussed below.

In the present analysis we used the correlation definition in Ref.~\cite{MCBias} that was derived from the mean-$p_T$ fluctuation quantity $\Delta\sigma^2_{p_T:n}$, developed by the STAR Collaboration~\cite{meanptpaper}. The resulting correlation quantity defines $\rho_{\rm se}$ and $\rho_{\rm me}$ such that the statistical bias caused by the multiplicity variation within a finite-width multiplicity bin is eliminated. 

In the present analysis charge-sign was determined and correlations for the four charge-pair combinations ($++$, $--$, $+-$, $-+$) were processed separately to ensure accurate efficiency and acceptance corrections. The bias-corrected, event-averaged number of like-sign (LS), same-event pairs and LS, mixed-event pairs, for arbitrary transverse rapidity bins $k,l$, are given by~\cite{MCBias}
\bea
\rho_{\rm se,kl}^{\pm\pm} & = & \frac{1}{\epsilon} \sum_{j=1}^{\epsilon} w_j^{{\rm se}\pm\pm} n_{j,kl}^{{\rm se}\pm\pm}
\label{Eq2} \\
\rho_{\rm me,kl}^{\pm\pm} & = & \frac{1}{\epsilon_{\rm mix}} \sum_{j\neq j^{\prime}} w^{{\rm me}\pm\pm} n_{j j^{\prime},kl}^{{\rm me}\pm\pm}.
\label{Eq3}
\eea
In Eqs.~(\ref{Eq2}) and (\ref{Eq3}) LS pairs ($++$,$--$) are indicated with superscripts, $\epsilon$ is the number of collision events in the centrality or multiplicity bin, index $j$ denotes a specific event, while in Eq.~(\ref{Eq3}) indices $j$ and $j^{\prime}$ denote arbitrary pairs of mixed-events where $\epsilon_{\rm mix}$ is the number of mixed-event permutations included in the multiplicity bin. The number of same-event, LS particle pairs from event $j$ in bin $(k,l)$ is given by quantity $n_{j,kl}^{{\rm se}\pm\pm}$ and similarly for mixed-event pairs where
\bea
n_{j j^{\prime},kl}^{{\rm me}\pm\pm} & = & n_{jk}^{\pm} n_{j^{\prime}l}^{\pm}
\label{Eq4}
\eea
and $n_{jk}^{\pm}$ is the single-particle count in bin $k$ for event $j$. The derivation in Ref.~\cite{MCBias} gives the event-wise weight factors
\bea
w_j^{{\rm se}\pm\pm} & = & \bar{N}^{\pm} / n_j^{\pm}
\label{Eq5} \\
w^{{\rm me}\pm\pm}   & = & (\bar{N}^{\pm} - 1)/\bar{N}^{\pm}
\label{Eq6}
\eea
where $n_j^{\pm}$ is the charged-particle multiplicity within the acceptance for event $j$ and $\bar{N}^{\pm}$ is the event ensemble average given by $\bar{N}^{\pm} = (1/\epsilon)\sum_j n_j^{\pm}$ within the event-multiplicity bin. All used events are required to have at least one LS pair.

For unlike-sign (US) pairs the results from Ref.~\cite{MCBias} give
\bea
\rho_{\rm se,kl}^{\pm\mp} & = & \frac{1}{\epsilon} \sum_{j=1}^{\epsilon} w_j^{{\rm se}\pm\mp} n_{j,kl}^{{\rm se}\pm\mp}
\label{Eq7} \\
\rho_{\rm me,kl}^{\pm\mp} & = & \frac{1}{\epsilon_{\rm mix}} \sum_{j^{\prime}\neq j^{\prime\prime}} w_{j^{\prime} j^{\prime\prime}}^{{\rm me}\pm\mp} n_{j^{\prime} j^{\prime\prime},kl}^{{\rm me}\pm\mp}.
\label{Eq8}
\eea
The event-wise weights are given by
\bea
w_j^{{\rm se}\pm\mp} & = & \sqrt{\frac{\bar{N}^+ \bar{N}^-}{n_j^+ n_j^-}}
\label{Eq9} \\
w_{j^{\prime} j^{\prime\prime}}^{{\rm me}\pm\mp} & = &
\sqrt{\frac{\bar{N}^{\pm} n^{\mp}_{j^{\prime}}} {\bar{N}^{\mp} n^{\pm}_{j^{\prime}}} } +
\sqrt{\frac{\bar{N}^{\mp} n^{\pm}_{j^{\prime\prime}}} {\bar{N}^{\pm} n^{\mp}_{j^{\prime\prime}}} } -
\frac{1}{\epsilon} \sum_{j=1}^{\epsilon}
\left[ \frac{n_j^+ n_j^-} {\bar{N}^+ \bar{N}^-} \right]^{1/2}. \nonumber \\
\label{Eq10}
\eea

Correlation quantities for each charged-pair combination are constructed as ratios defined in Eq.~(\ref{Eq2}) and are given by
\bea
\left(\frac{\Delta\rho}{\rho_{\rm me}}\right)^{ab}_{kl} & \equiv &
\frac{ \rho_{{\rm se},kl}^{ab} - \rho_{{\rm me},kl}^{ab}}{\rho_{{\rm me},kl}^{ab}},
\label{Eq11}
\eea
where superscripts $(a,b)$ denote charge-sign combinations. LS, US, all-charges or charge-independent (CI), and charge-difference or charge-dependent (CD) combinations are constructed from the ratios in Eq.~(\ref{Eq11}) for final reporting of results. The four combinations are
\bea
\left( \frac{\Delta\rho}{\rho_{\rm me}} \right)^{\rm LS}_{kl} & = &
\frac{1}{2} \sum_{ab=++,--} \left( \frac{\Delta\rho}{\rho_{\rm me}} \right)^{ab}_{kl}
\label{Eq12} \\
\left( \frac{\Delta\rho}{\rho_{\rm me}} \right)^{\rm US}_{kl} & = &
\frac{1}{2} \sum_{ab=+-,-+} \left( \frac{\Delta\rho}{\rho_{\rm me}} \right)^{ab}_{kl}
\label{Eq13} \\
\left( \frac{\Delta\rho}{\rho_{\rm me}} \right)^{\rm CI}_{kl} & = &
  \frac{1}{2} \left( \frac{\Delta\rho}{\rho_{\rm me}} \right)^{\rm LS}_{kl} +
       \frac{1}{2} \left( \frac{\Delta\rho}{\rho_{\rm me}} \right)^{\rm US}_{kl}
\label{Eq14} \\
\left( \frac{\Delta\rho}{\rho_{\rm me}} \right)^{\rm CD}_{kl} & = &
 \frac{1}{2} \left( \frac{\Delta\rho}{\rho_{\rm me}} \right)^{\rm LS}_{kl} -
\frac{1}{2} \left( \frac{\Delta\rho}{\rho_{\rm me}} \right)^{\rm US}_{kl}.
\label{Eq15}
\eea
Acceptance and single-particle reconstruction inefficiency effects cancel in the ratios in Eqs.~(\ref{Eq11})-(\ref{Eq15}) since these effects are present in both the same- and mixed-event quantities. Two-particle reconstruction inefficiencies do not cancel and require an additional correction procedure (see Sec.~\ref{SecIII}).

The prefactors are calculated using analytic representations of the efficiency- and acceptance-corrected charged-particle distributions on transverse rapidity. In addition, the prefactors must account for the number of LS and US particle pairs, as well as the number of near-side and away-side pairs, respectively. The final CI, all-azimuth normalized correlation quantity used in this analysis is defined by
\bea
\left( \frac{\Delta\rho}{\sqrt{\rho_{\rm chrg}}} \right)^{\rm CI}_{kl} & \equiv &
{\cal P}^{\rm CI,All}_{kl} \left( \frac{\Delta\rho}{\rho_{\rm me}} \right)^{\rm CI}_{kl}
\label{Eq16}
\eea
where ${\cal P}^{\rm CI,All}_{kl}$ is the CI, all-azimuth prefactor. Prefactors for the other charge-pair combinations and relative azimuthal angle pair projections are obtained by scaling the above prefactor according to the average number of pairs. Details of the efficiency-corrected particle distributions, and the scale factors for each charge combination and azimuthal angle selection, for all prefactors used in this analysis, are given in Appendix~\ref{Appendix-A}.

\section{Data}
\label{SecIII}

Data for this analysis were taken with the STAR detector~\cite{star} during the 2004 RHIC Run (Run~4) as described in Ref.~\cite{axialCI}. Minimum-bias triggered events for Au+Au  collisions at energy $\sqrt{s_{\rm NN}}$ = 200~GeV were obtained by requiring a coincidence of two Zero-Degree Calorimeters (ZDCs) and a minimum number of charged-particle hits in the Central Trigger Barrel scintillator material~\cite{startrig}. Charged-particle measurements with the Time Projection Chamber (TPC)~\cite{STARTPC} and event triggering are described in Ref.~\cite{star}. Charged particle trajectories were measured in a uniform 0.5~T magnetic field which was alternately oriented parallel and anti-parallel to the beam axis to evaluate systematic tracking errors. Primary vertices (PV) along the beam axis ($z$-axis) were reconstructed using TPC tracks and were required to be within 25~cm of the geometrical center of the TPC. The data accepted for this analysis included 9.5 million events. The available data sample was sufficient to measure the correlation structures of interest. The focus of the present measurements are the correlations associated with non-identified charged particles within the low-to-intermediate $p_T$ range corresponding to the bulk of the produced particles from the most peripheral (most similar to the p+p limit) to most-central collisions. 

Accepted particle trajectories (tracks) were required to be within the optimum TPC acceptance, defined by $p_T>0.15$~GeV/$c$, $|\eta| < 1.0$ and $-\pi \leq \phi \leq \pi$. All accepted tracks used in the analysis were required to have at least 20 (out of a possible 45) reconstructed space points in the TPC, a ratio of the number of found space points to the maximum number expected $>$ 0.52 (to eliminate split tracks), a least-squares fitted $\chi^{2}/{\rm NDF} < 3$ (number of independent degrees of freedom $-$ NDF), and a distance of closest approach (DCA) of the projected trajectory (helix) to the primary collision vertex $<$~3~cm. Accepted particles included true primary hadrons from the collision plus approximately 12\% background contamination~\cite{starspec200,PRC79}
from weak decays and interactions within the detector material. Backgrounds from photon conversion to electron-positron pairs were reduced by excluding particles with $dE/dx$ (ionization energy loss in the TPC gas) within $1.5\sigma$ of that expected for electrons in the momentum ranges $0.2 < p < 0.45$~GeV/$c$ and $0.7 < p < 0.8$~GeV/$c$~\cite{axialCI}. Particle identification was not implemented, but charge sign was determined via the direction of track curvature in the magnetic field~\cite{star}. Corrections for two-track reconstruction inefficiencies were applied to the $\rho_{\rm se}/\rho_{\rm me}$ ratios using two-track separation distance cuts as described in Appendix~C of Ref.~\cite{axialCI}. Further details of track definitions, efficiencies, and quality cuts are described in Refs.~\cite{starspec200,mikeThesis,LizThesis,PrabhatThesis}.

Event pileup is caused by untriggered events in beam-beam bunch crossings that occur within the TPC drift time (35$\mu$s) before or after the bunch crossing that contains the triggered event. These out-of-time collisions produce particle trajectories in the TPC which can be erroneously reconstructed as the triggered event, or which contaminate the particle trajectories reconstructed from the triggered event. Although the pileup rate in Run~4 was typically less than 0.4\%, this level of contamination was shown to produce significant artifacts in the angular correlations~\cite{axialCI}. The pileup filter and correction procedure described in Appendix~D of Ref.~\cite{axialCI} was applied in the present analysis. Pileup effects in the transverse-rapidity correlations are much less significant than they are in the angular correlations \cite{axialCI} (see Sec.~\ref{SecV}). 

The minimum-bias event sample comprised 0-93\% of the total reaction cross section and was divided into eleven centrality bins, using the event-wise number of accepted TPC tracks (particles) with $|\eta| \leq 1$ and $p_T \geq 0.15$~GeV/$c$ as described in Ref.~\cite{axialCI}. The measured multiplicity frequency distribution for the Run~4 minimum-bias data was approximately the same as in the 2002 data run which was analyzed in Ref.~\cite{axialCI}, where centrality bins based on accepted track multiplicity cuts were determined. Those same multiplicity cuts were used in the present analysis to facilitate direct comparison with the angular correlations.\footnote{Centrality was based on multiplicities within $|\eta| \leq 1$ in order to avoid significant artifacts in the angular correlations along the $\Delta\eta$ direction and within the range $|\Delta\eta| \leq 2$~\cite{axialCI}.} Additional corrections due to small (few percent) variations in the TPC tracking efficiency as functions of PV position and run-time luminosity were negligible and therefore not corrected for. 

Correlations were calculated for each centrality by grouping events based on the PV position along the beam line and on event-wise multiplicity. The former was done in order to suppress systematic error caused by particle-pair event-mixing between collisions for which track reconstruction acceptance and efficiency differ with PV position in the TPC. The multiplicity grouping within a centrality was required to suppress systematic effects caused by overall slope changes and other shape variations in the single-particle $p_T$ distributions within broad centrality ranges~\cite{phenom}. In addition, the event-mixing procedure was only performed with events taken within the same data acquisition run (typically 30-60 minutes) where detector performance remained relatively stable. 

From previous correlation analyses of 200~GeV Au+Au collision data using the STAR TPC tracking detector~\cite{axialCI,STARptscale1,STARptscale2,Kettlerptdependent}, it was determined that PV positions within 5~cm and event-wise multiplicities within 50 are sufficient to achieve stable correlations. The 50~cm PV position range was therefore divided into 10 uniform sub-bins and the centrality range was divided into 22 multiplicity sub-bins~\cite{mikeThesis}. The PV position sub-binning was only required for the three most-central bins covering the cross-section range from 0-18\%. Ratios $\Delta\rho/\rho_{\rm me}$ in each PV and multiplicity sub-bin, and for each data acquisition run, were combined over the entire data volume using total pair-number weighted averages to produce the final correlations in the eleven centrality bins. Pre-factors were applied to the final weighted averages of ratios.


The $(y_{T1},y_{T2})$ bins were filled with all charged-particle pairs within the full TPC angular acceptance that fall within selected ranges of relative azimuth where the $|\Delta\phi|$ ranges include $ \leq \pi/2$ [near-side (NS)], $> \pi/2$ [away-side (AS)], and $0 \leq |\Delta\phi| \leq \pi$ (all azimuth angles). Pair weights correcting for finite $\eta$ acceptance were not included. Results are presented for LS, US, CI, and CD combinations. The present dataset includes 132 correlation distributions on $(y_{T1},y_{T2})$. The transverse rapidity range is $y_T \in [1.0,4.5]$, corresponding to $p_T \in [0.16,6.3]$~GeV/$c$. The $(y_{T1},y_{T2})$ space was uniformly binned into a 25$\times$25 grid corresponding to bin coordinates $k,l$ introduced in Sec.~\ref{SecII}.

For each same-event pair and mixed-event pair both permutations were counted in filling the histograms, resulting in symmetric correlations, i.e. $\Delta\rho(y_{T1},y_{T2}) = \Delta\rho(y_{T2},y_{T1})$ and $\rho_{\rm me}(y_{T1},y_{T2}) = \rho_{\rm me}(y_{T2},y_{T1})$, or equivalently $\Delta\rho_{k,l} = \Delta\rho_{l,k}$ and $\rho_{{\rm me},kl} = \rho_{{\rm me},lk}$. Statistical errors in diagonal bins ($y_{T1} = y_{T2}$ or $k=l$) were computed according to the total number of unique particle pairs, for both same-events and mixed-events, in each bin~\cite{ReidNIM,RayNIM}. Statistical errors were similarly computed in off-diagonal bins with $y_{T1} > y_{T2}$ and then applied to the corresponding bins with $y_{T1} < y_{T2}$.
The mixing algorithm used here and elsewhere results in reduced statistical noise, as explained in Refs.~\cite{ReidNIM} and as applied to the present event-mixing method in Ref.~\cite{RayNIM}. In the present analysis, mixed-event particle-pairs were constructed using all accepted particles from one event with all accepted particles in the next two events in the event-list. This process was iterated through all events in each PV and multiplicity bin.

The typical statistical errors for the CI, all-azimuth correlations are approximately 5\% of the peak amplitude in the correlation structure near $(y_{T1},y_{T2}) \approx (3,3)$. The magnitudes of the statistical errors are approximately the same for the $\Delta\phi$ and charge-pair projections when scaled by the corresponding prefactors. Similarly, the magnitudes of the errors for CD correlations are approximately the same as those for the corresponding CI correlations. The statistical errors increase in magnitude toward larger $y_T$ and near the off-diagonal corners. Due to symmetrization of the correlation data, the statistical errors in diagonal $y_{T1}=y_{T2}$ bins are approximately $\sqrt{2}$ times larger than those in neighboring, off-diagonal bins.

\begin{figure*}[t]
\includegraphics[keepaspectratio,width=6.5in]{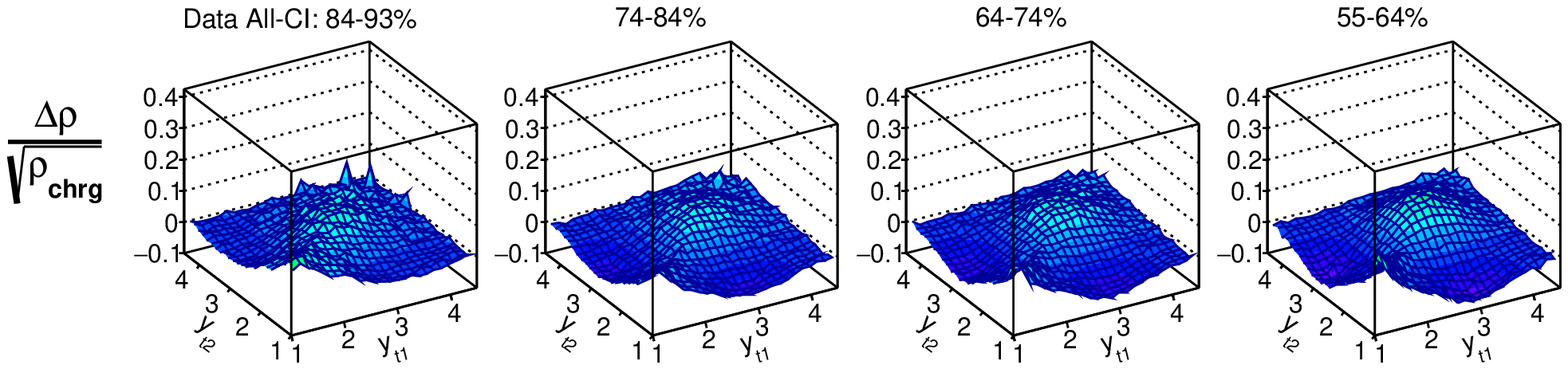}
\includegraphics[keepaspectratio,width=6.5in]{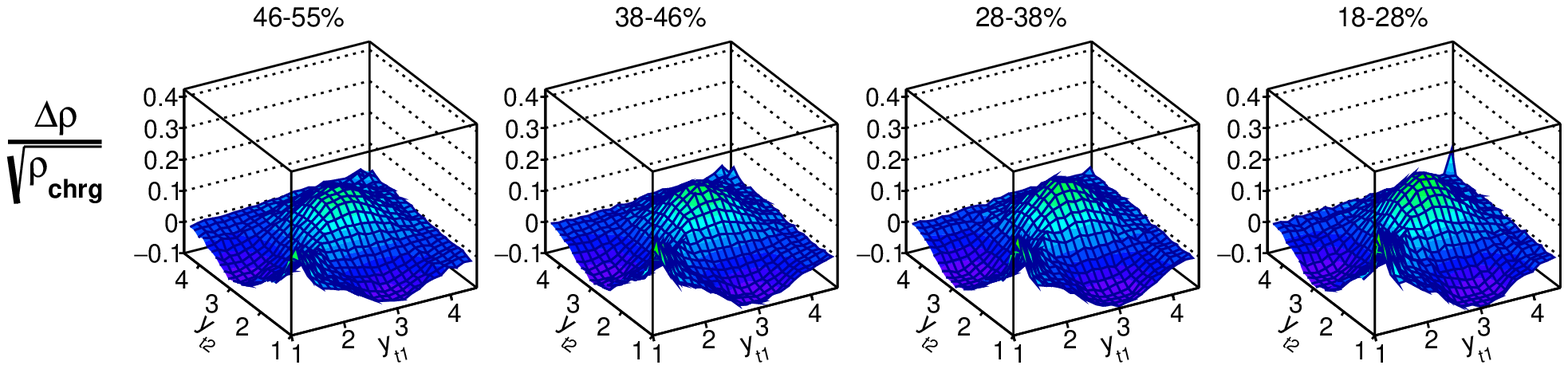}
\includegraphics[keepaspectratio,width=4.875in]{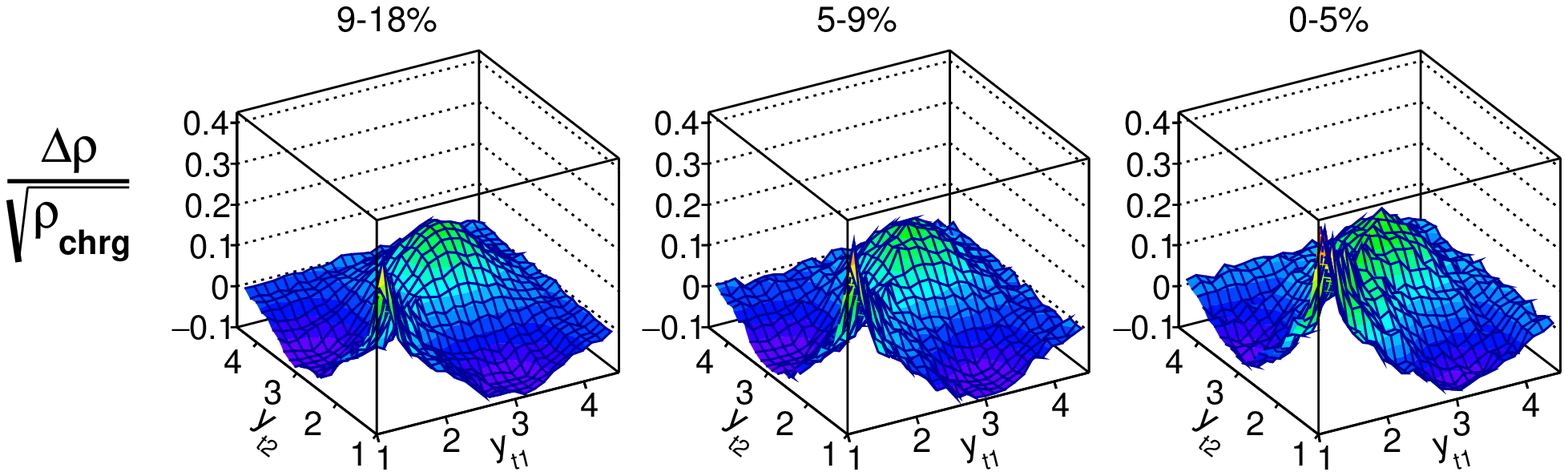}
\caption{\label{Fig1}
Perspective views of two-dimensional correlations $\Delta\rho/\sqrt{\rho_{\rm chrg}}$ on coordinates $(y_{T1},y_{T2})$ (pion mass assumed) for minimum-bias Au+Au collisions at $\sqrt{s_{\rm NN}}$ = 200 GeV using all charged particle pairs and including all relative azimuthal angles $\Delta\phi$ from $-\pi$ to $\pi$, as discussed in the text. Centrality ranges are indicated for each panel in percent of total hadronic reaction cross section.}
\end{figure*}

\begin{figure*}[t]
\includegraphics[keepaspectratio,width=6.5in]{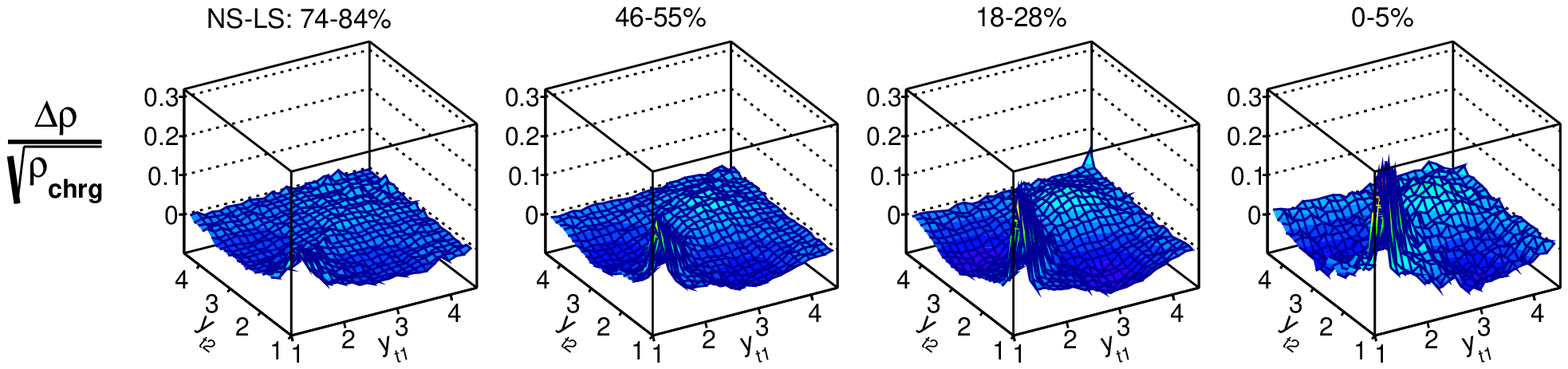}
\includegraphics[keepaspectratio,width=6.5in]{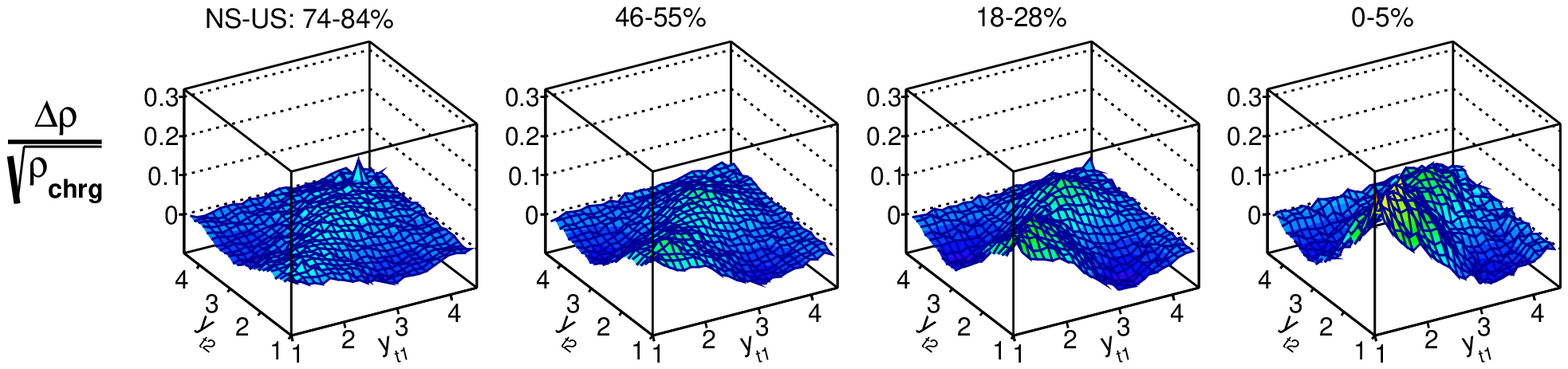}
\includegraphics[keepaspectratio,width=6.5in]{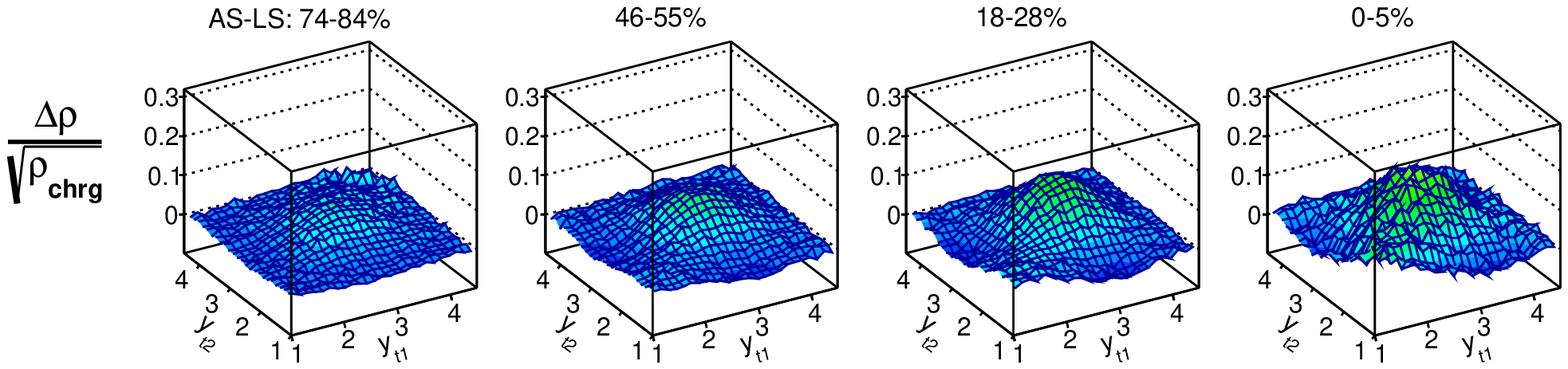}
\includegraphics[keepaspectratio,width=6.5in]{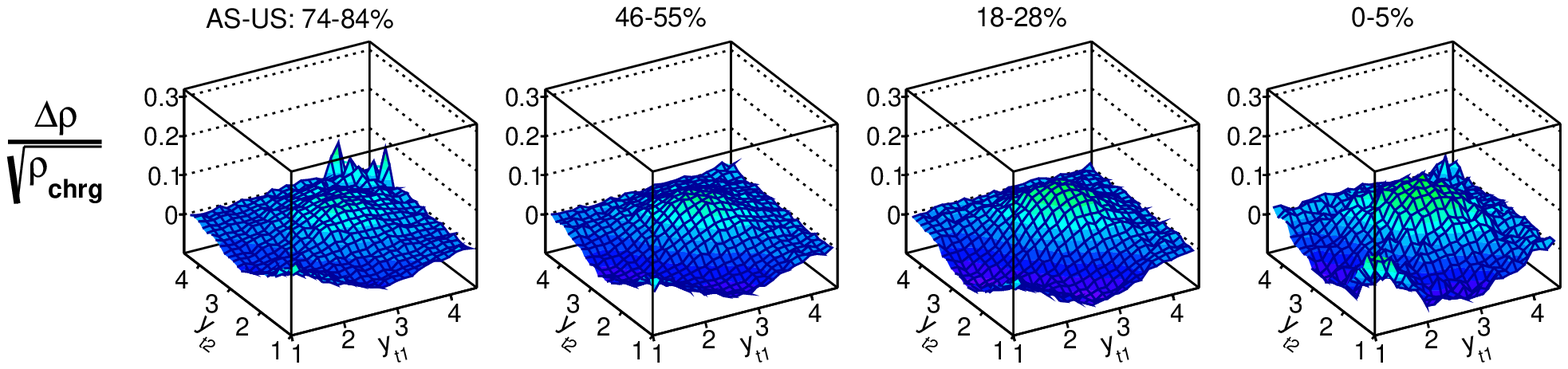}
\caption{\label{Fig2}
Perspective views of two-dimensional correlations $\Delta\rho/\sqrt{\rho_{\rm chrg}}$ on coordinates $(y_{T1},y_{T2})$ for Au+Au collisions at $\sqrt{s_{\rm NN}}$ = 200 GeV as discussed in the text. The first two rows correspond to charged particle pairs with relative azimuth $|\Delta\phi| \leq \pi/2$ (near-side).  The bottom two rows correspond to charged particle pairs with relative azimuth $\pi \geq |\Delta\phi| >\pi/2$ (away-side). The first and third rows are for LS pairs and the second and fourth rows are for US pairs.  Centrality varies in each row of panels from left-to-right from peripheral to most-central corresponding to total cross-section fractions 74-84\%, 46-55\%, 18-28\%, and 0-5\%, respectively.}
\end{figure*}

\begin{figure*}[t]
\includegraphics[keepaspectratio,width=6.5in]{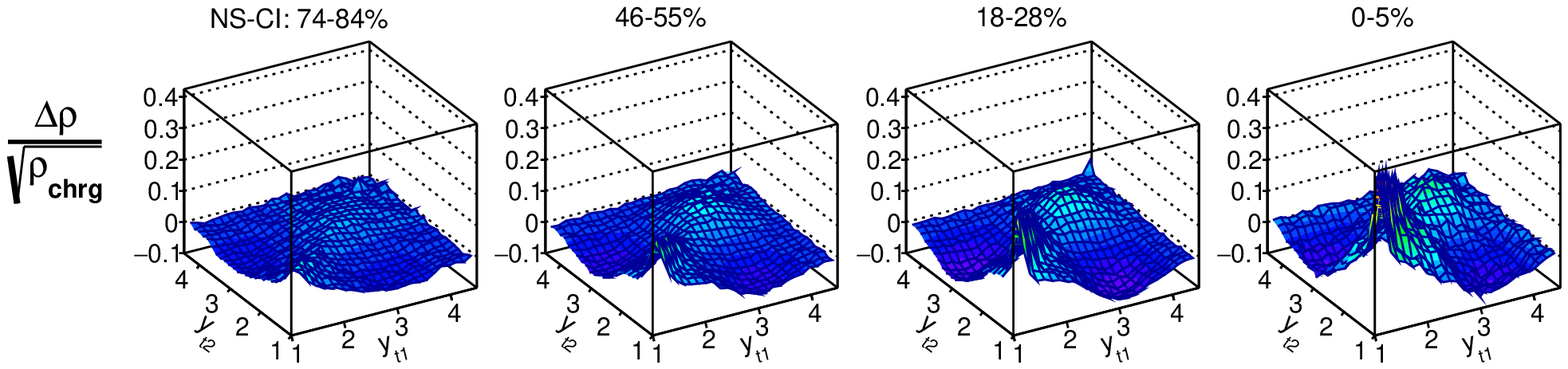}
\includegraphics[keepaspectratio,width=6.5in]{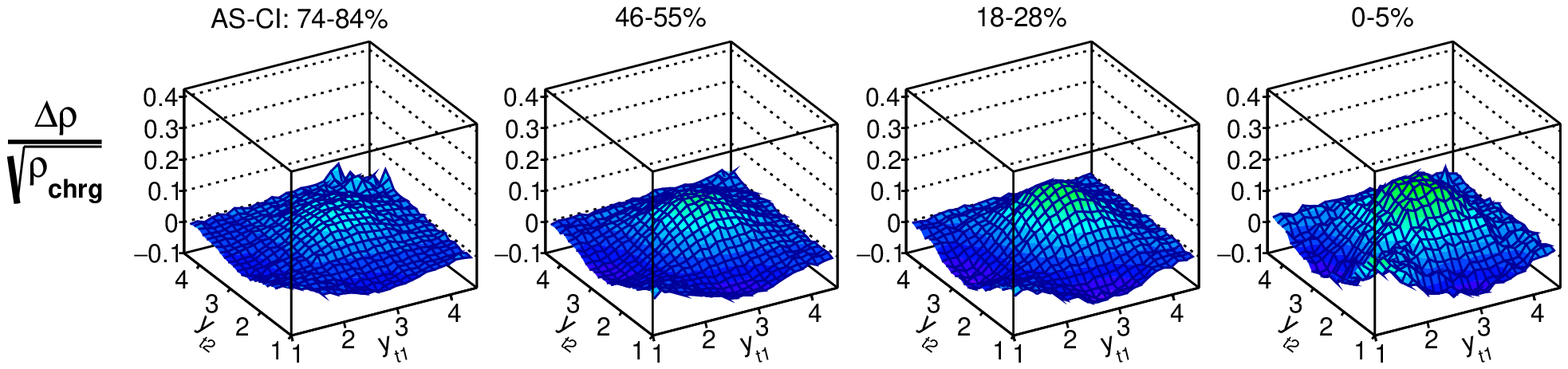}
\includegraphics[keepaspectratio,width=6.5in]{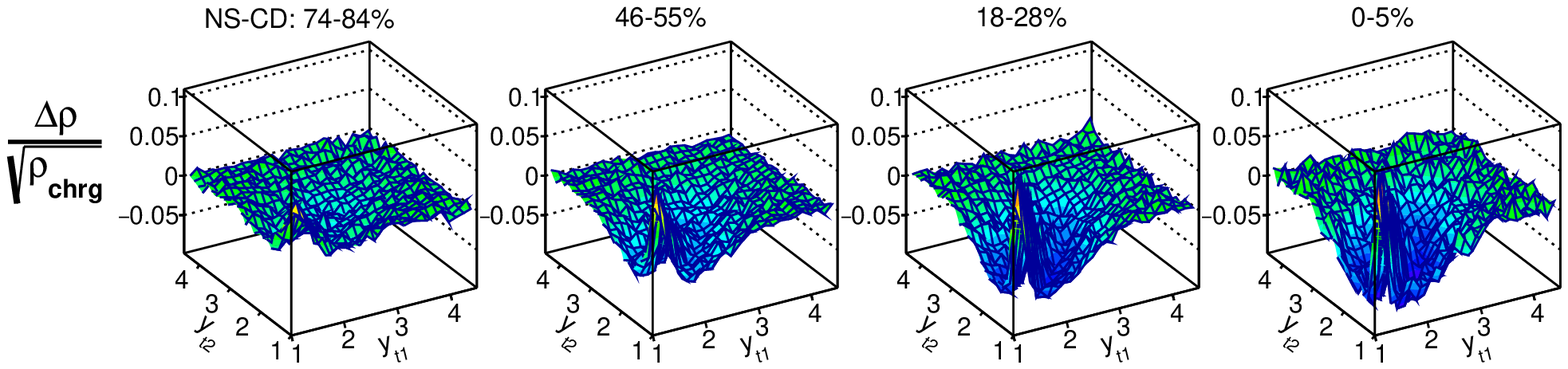}
\includegraphics[keepaspectratio,width=6.5in]{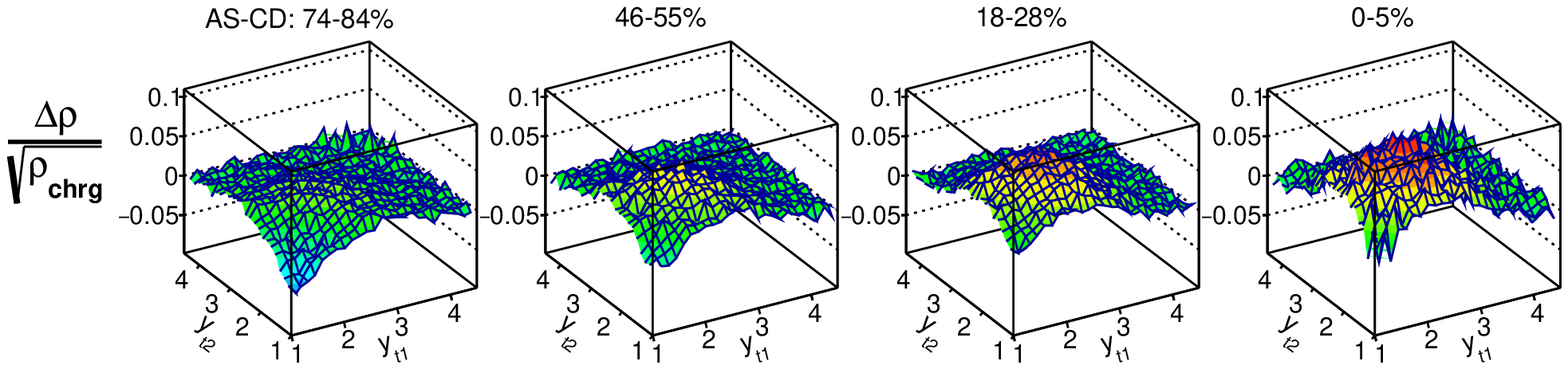}
\caption{\label{Fig3}
	Same as Fig.~\ref{Fig2} except for the sum (CI) (upper two rows) and differences (CD) (lower two rows) between LS and US charged-pairs for NS relative azimuth (first and third rows) and AS (second and fourth rows) as discussed in the text. Centrality varies in each row of panels from left-to-right for total cross-section fractions 74-84\%, 46-55\%, 18-28\%, and 0-5\%, respectively.}
\end{figure*}

\begin{figure*}[t]
\includegraphics[keepaspectratio,width=5.5in]{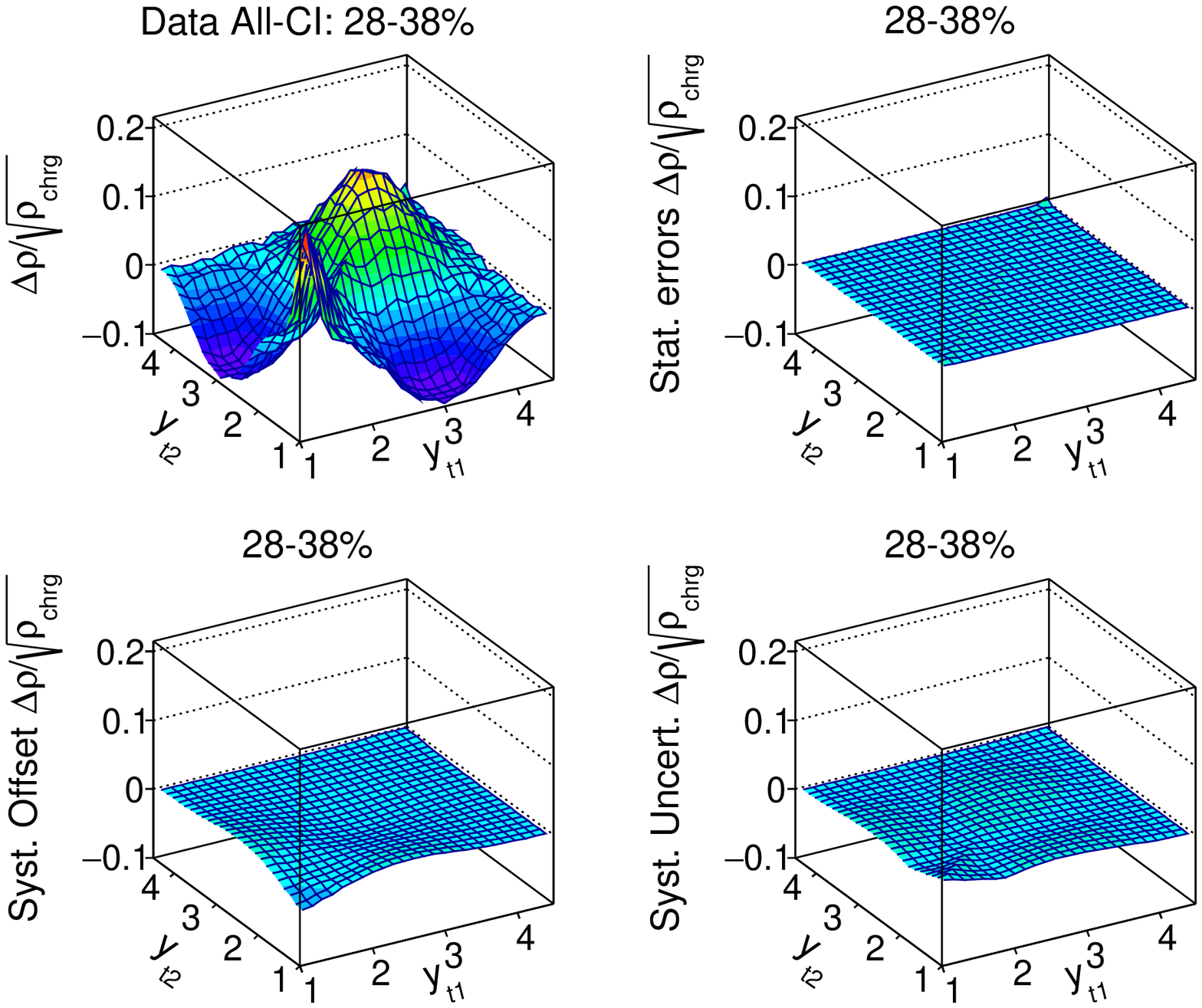}
\caption{\label{Fig4}
Statistical errors, systematic offsets, and systematic uncertainties in comparison with the data for each $(y_{T1},y_{T2})$ bin for the CI, all-azimuth correlations for the 28-38\% centrality bin. A common scale is used to emphasize the relative magnitudes of the correlations and the errors.}
\end{figure*}

\section{Correlation measurements}
\label{SecIV}

In this section, representative examples of our correlation measurements are presented and the prominent features are noted and discussed. Comparisons with theoretical predictions are presented in Sec.~\ref{SecVI}. Possible physical interpretations of the correlation structures presented in this section are discussed in Sec.~\ref{SecVII}.

Perspective views of the CI, all-azimuth correlations are shown in Fig.~\ref{Fig1} for the eleven centrality bins. The structural features include a monotonically increasing peak along the main diagonal near $y_T \approx 3$ (pion mass assumption) corresponding to $p_T \approx 1.4$~GeV/$c$, a pronounced saddle shape, and a ridge along the main diagonal at lower $y_T$ which can be attributed to Bose-Einstein quantum-correlations~\cite{HBT}. In general the observed correlation structures smoothly increase with centrality. The amplitudes of the maxima near $(y_{T1},y_{T2}) \approx (3,3)$ and the saddle-shape minima near $(y_{T1},y_{T2}) \approx (3,1)$ ($y_T$ = 1 corresponds to $p_T$ = 0.16~GeV/$c$) vary smoothly with centrality, generally increasing in amplitude from peripheral to central collisions. The positions of the maxima and minima are generally stable, but with some, modest variation with centrality. All of these features are significant with respect to statistical and systematic uncertainties (see Sec.~\ref{SecV}).

Perspective views of CD correlations on transverse rapidity with either near-side or away-side relative azimuthal angles for four centrality bins from peripheral to most-central are shown in Fig.~\ref{Fig2}. The four columns of panels display the centrality dependence for the 74-84\%, 46-55\%, 18-28\% and 0-5\% bins as indicated by the labels at the top of the figure. The rows of panels from upper to lower correspond to near-side, like-sign pairs (NS-LS), near-side, unlike-sign pairs (NS-US), away-side, like-sign pairs (AS-LS), and away-side, unlike-sign pairs (AS-US), respectively.

For the NS-LS correlations the sharp, positive peaks along the main diagonal are produced by Bose-Einstein quantum correlations~\cite{HBT}, predominantly among identical, charged pions. Those features increase in amplitude with centrality according to the total number of identical particle pairs in the emission region~\cite{HBT}. An overall saddle-shaped structure is also apparent which increases moderately in amplitude with centrality. The lower $y_T$ saddle feature is partially obscured by the quantum correlation structure. The peak along the main diagonal, whose maximum is near $y_T \approx 3$, also increases monotonically with centrality. The shape of this peak in $(y_{T1},y_{T2})$ space is approximately symmetric with respect to the widths along the sum ($y_{T\Sigma} = y_{T1} + y_{T2}$) and difference ($y_{T\Delta} = y_{T1} - y_{T2}$) directions.

For the NS-US correlations, a double-peaked structure appears along the main diagonal with one maxima at $y_T \approx$ 2.0 to 2.5 ($p_T \approx$ 0.5 to 0.85~GeV/$c$) and the second near $y_T \approx 3$. Both of these peak structures monotonically increase in amplitude with centrality. The peaked structure at lower $y_T$ is most pronounced in the NS-US projection. The peak at larger $y_T$ is asymmetric where the width along $y_{T\Sigma}$ is larger than the width along $y_{T\Delta}$. The magnitudes of the saddle-shape minima increase from 0.03 to 0.07 with centrality.
Conversion electron-positron pairs that pass the cuts produce angular correlations with small opening angles~\cite{axialCI} and are therefore a potential source of contamination in the NS-US projection. 
Simulations, discussed in Sec.~\ref{SecV}, show that conversion electron pair contamination is very small relative to the NS-US correlations and mainly contributes along the lower-momentum edges of the $(y_{T1},y_{T2})$ domain for $y_{T} < 2.5$. This contamination is much smaller than the two-peaked correlation structure of interest here.


The AS-LS correlations display an overall saddle shape with a monotonically increasing peak along the main diagonal with maximum at $y_T \approx  3$. The 2D peak widths along the $y_{T\Sigma}$ and $y_{T\Delta}$ directions are approximately equal. The low $y_T$ peak at $(y_{T1},y_{T2}) \approx (1,1)$ also increases with centrality as does the depth of the saddle minimum. 

The AS-US correlation structures are similar to those of the AS-LS. However, the $(y_{T1},y_{T2}) \approx$ (3,3) peak widths are asymmetric, being elongated in the difference direction along $y_{T\Delta}$ relative to the sum direction. Also, the low $y_T$ peak displays a different centrality dependence. For peripheral collisions this structure appears to subside with increasing centrality, producing a minimum along the $y_{T1} = y_{T2}$ diagonal which merges with the saddle-shape minimum. For more-central collisions the peak at $(y_{T1},y_{T2}) \approx (1,1)$ partially re-emerges. 
Quantum correlations and conversion electron contamination do not contribute to these correlations which require $|\Delta\phi| > \pi/2$.

In Fig.~\ref{Fig3} the charge-independent (LS + US) NS and AS correlations are shown in the first two rows of panels, respectively. The peaks near $(y_{T1},y_{T2}) \approx (3,3)$ for both the NS and AS correlations increase monotonically with centrality, the AS amplitudes being larger than the corresponding NS amplitudes. Both sets of peaked structures are asymmetric; those on the NS are elongated along the $y_{T\Sigma}$ direction while those on the AS are elongated along the $y_{T\Delta}$ direction. The HBT correlations are prominent in the NS correlations and may partially obscure a low $y_T$ saddle-shape peak at $(y_{T1},y_{T2}) \approx (1,1)$.

In the third and fourth rows of panels in Fig.~\ref{Fig3} the charge-dependent (LS $-$ US), NS and AS correlations are shown. The magnified $z$-axis scale causes the statistical fluctuations (individual spikes) to be more pronounced than in the upper two rows of panels for the CI correlations. For the correlation peaks near $(y_{T1},y_{T2}) \approx (3,3)$, those on the NS associated with US pairs are systematically larger than those from LS pairs. This leads to negative CD correlations in this larger $y_T$ range for NS pairs. At lower $y_T$ the US peak is stronger than a possible LS peak (other than HBT correlations) producing deep minima along the main diagonal. The NS negative CD correlations are evident from $y_T$ = 1.0 to about 3.5 ($p_T$ = 0.16 to 2.3~GeV/$c$) and monotonically deepen with centrality. The negative correlations are elongated in the sum direction $(y_{T\Sigma})$ relative to their widths along $(y_{T\Delta})$.

For AS-CD correlations in the last row of Fig.~\ref{Fig3} the approximate equality of LS and US peak amplitudes near $(y_{T1},y_{T2}) \approx (3,3)$ lead to approximately zero CD correlations in this region. The subsidence of the AS-US correlations at lower $y_T$ (bottom row of Fig.~\ref{Fig2}) leads to positive CD correlations at lower $y_T$ (bottom row of Fig.~\ref{Fig3}) producing a pronounced peak at $y_{T1} = y_{T2} \approx 2.2$ ($p_T \approx$ 0.62~GeV/$c$) which monotonically increases with centrality.


\section{Systematic uncertainties}
\label{SecV}

Systematic uncertainties in the correlation measurements arise from secondary particle contamination, photon conversion to correlated electron-positron pairs in the detector material, event pileup in the TPC, ambiguities in the two-track reconstruction inefficiency corrections, relative separation distance cuts between mixed-event PV locations, event multiplicity differences for event-mixing, PV position and beam luminosity dependent track reconstruction inefficiency, systematic bias in the correlation measure quantity itself~\cite{MCBias}, and uncertainties in the charged-particle multiplicity. Other sources of systematic uncertainty identified for the Au + Au Run~4 data and discussed in Ref.~\cite{axialCI} were estimated to be negligible for the present correlation measurements and were therefore not included in the systematic uncertainties. 

The primary particle sample for the STAR Run~4 Au+Au 200~GeV collision data includes approximately 12\% contamination from weak-decay daughter particles and from pions and protons produced in the detector material between the collision vertex and the TPC tracking volume~\cite{PRC79}. Reducing the maximum allowed DCA to the primary vertex from 3~cm to 1~cm reduced this contamination, but also reduced the primary particle yield, especially at lower $p_T$. Reducing the primary particle yield at lower $p_T$ distorts the true correlations, confounding efforts to identify the effects of the secondary particles. Simulations were used to estimate the systematic uncertainties, where a model of the secondary particle $p_T$ spectra~\cite{PRC79} was used in which the amplitude and overall slope were allowed to independently fluctuate from event-to-event. Details of the calculations are given in Appendix~\ref{Appendix-C}. Poisson fluctuations in the event-wise secondary particle yield produced significant uncertainties in the correlations, mainly at lower $y_T$. Fluctuations in the slope of the secondary particle $p_T$ spectra produced much smaller effects.

Photon conversions to $e^+e^-$ pairs in the detector material were estimated using the Monte Carlo simulation described in Ref.~\cite{LizThesis}. For the Run~4 STAR detector configuration those materials included the beam pipe, the Silicon Vertex Tracker (SVT)~\cite{SVT}, and the TPC inner field-cage. In the simulation, a realistic $\pi^0$ $p_T$ spectrum was assumed, where random pion decays, $\pi^0 \rightarrow \gamma + \gamma$, were included ($\eta  \rightarrow \gamma + \gamma$ decays were not included), followed by $\gamma + A \rightarrow e^+ + e^- + A^{\star}$ conversion processes in the detector material calculated using the Bethe-Heitler equation~\cite{PDG}. The average yield of correlated $e^+e^-$ pairs was estimated by normalizing to the volume of the sharp 2D exponential angular correlation at $(\Delta\eta,\Delta\phi) = (0,0)$ reported in Ref.~\cite{axialCI} for US charged-particle pairs in 200~GeV Au+Au collisions~\cite{LizThesis}. The $e^+e^-$ pairs are primarily produced with $y_{T_\Sigma} < 3.5$. Correlations on transverse rapidity between the $e^+e^-$ pairs of each $\gamma$-conversion process are generated by pair-production dynamics and are proportional to the average number of $\gamma$-conversions per event. This background contribution could, in principle, be subtracted from the measured correlations. However, this estimate is considered to be quite uncertain due to the potentially large contributions from final-state Coulomb interactions. The estimated $\gamma$-conversion contribution is therefore considered an uncertainty where one-half of the estimated correlation contribution is assumed to be a systematic offset and $\pm$ one-half is the systematic uncertainty. The uncertainties range from about 0.002 to 0.003 at lower $y_T$ from peripheral to most-central collisions, respectively. These contributions are small relative to those from other secondary particles. 

Most pileup contamination was removed and corrected using the procedure described in Appendix~D of Ref.~\cite{axialCI}. However, it is likely that some residual contamination remains. This was estimated in Ref.~\cite{axialCI} to be about $\pm$10\% of the full pileup contribution for these data. To estimate this effect, the CI, all-azimuth correlations were constructed without the pileup filter and correction procedure. Those correlations were subtracted from the final, pileup corrected correlation data. One-tenth of the net difference was used to estimate the systematic uncertainty that was approximated with a 2D Gaussian given by $\pm A \exp [-((y_{T1} - y_{T0})^2 + (y_{T2} - y_{T0})^2)/2\sigma^2]$. From the mid-centrality bin 64-74\% to the 18-28\% bin, amplitude $A$ = 0.00055, 0.0013, 0.0023, 0.0026, 0.0017, 0.00037; peak position $y_{T0}$ varies from 2.0 to 2.3; and width $\sigma$ varies from about 1.0 to 0.5, respectively. Pileup effects are negligible for the other more-peripheral and more-central bins.

Particle pair reconstruction inefficiencies~\cite{PRC73-64907,axialCI} were approximately corrected using two-track separation distance cuts in the TPC~\cite{STARHBT1,mikeThesis,LizThesis,axialCI}. Residual effects may continue to exist and were estimated by comparing the $(y_{T1},y_{T2})$ correlations computed assuming different separation distance averaging methods and/or cut values. Bin-wise differences provided an estimate of the systematic uncertainties. These differences could also be approximated with a 2D Gaussian where amplitude $A$ varied from  $\pm 0.0013$ to $\pm 0.0009$, $y_{T0}$ varied from 2.3 to 2.5, and $\sigma$ varied from 0.5 to 0.7 for centrality bins from 46-55\% to 18-28\%, respectively. This uncertainty was negligible for the other centrality bins.

For the event-mixing procedure to be accurate, the events being mixed must be similar as explained in Sec.~\ref{SecIII}. Previous analyses~\cite{mikeThesis,LizThesis,axialCI} showed that for the STAR Run~4 Au+Au 200~GeV collision data the allowed event-mixing multiplicity range, with $|\eta|<1$ acceptance, must be $\leq$~50, and the primary vertex positions along the beam axis, for event-mixing, must be within 5~cm. The present correlations remained stable, i.e. no systematic effects, when the allowed multiplicity range was reduced below 50. Restricting the mixed-event pair PV relative positions to be $<$ 5~cm had no significant effect in the three most-central bins from 0-18\%, but did produce a small net increase in the correlations at large $y_T$ in the seven centrality bins from 18-84\%. This systematic increase was approximated by an exponential function $A \exp [(y_{T_\Sigma} - 8.5)/0.2]$ for $y_{T_\Sigma} \leq 8.5$, and constant $A$ for $y_{T_\Sigma} > 8.5$. This entire effect is considered an uncertainty and is represented in each bin with an offset and $(\pm)$ uncertainty, where both equal one-half the value of the preceding function. The magnitudes vary from 0.004 to 0.0035 from peripheral to central (84\% to 18\%) collisions. This uncertainty is mainly confined to the upper $(y_{T1},y_{T2})$ corner with $y_T > 4$ ($p_T >$ 3.8~GeV/$c$). 

Track reconstruction efficiency in the STAR TPC is reduced when the PV position shifts along the beam axis away from the geometrical center of the detector. For the present data the collision vertices were accepted within $\pm25$~cm of the center of the TPC. Tracking efficiency is also reduced when beam luminosity increases, e.g., at the beginning of each beam fill in the collider, due to the increased space-point hit density in the TPC gas volume. Coincidence rates in the ZDCs~\cite{startrig}, a measure of luminosity, varied from about 10~kHz at the beginning of a beam fill in the collider down to about 1~kHz at the end of the fill. Tracking efficiency decreases an additional 4.5\% for collisions occurring at $\pm$25~cm, and 3\% when coincidence rates reach as high as 10~kHz. These position and luminosity dependent tracking efficiency effects were not used to correct event-wise multiplicity, resulting in small, systematic shifts in the centrality assignments for each event. Because the correlations systematically vary with centrality, these systematic shifts introduce a systematic error. This effect was studied with the Monte Carlo simulation described in Ref.~\cite{MCBias} and shown to be negligible. The correlation amplitudes were affected by 0.0003 or less in more-peripheral collisions and by 0.0001 or less in more-central collisions.

In Fig.~5 of Ref.~\cite{MCBias}, the systematic variation in the overall slope of the single particle $p_T$ spectrum with respect to event multiplicity, occurring within an event-mixing group, introduces a systematic bias in the measured $(y_{T1},y_{T2})$ correlations. One-half of this bias in each bin was assumed to be a systematic offset and $\pm 1/2$ of the bias was taken to be the uncertainty.

Finally, the prefactor includes systematic uncertainties that are dominated by the normalization uncertainty in the measured charged particle multiplicity $dN_{\rm ch}/d\eta$~\cite{PRC79}. The Au+Au 200~GeV multiplicities are consistent with $p_T$ spectra reported by STAR~\cite{STARspectra}. The systematic uncertainties range from $\pm$10\% in more-peripheral to $\pm$7\% in most-central collisions. 

In summary, the dominant systematic uncertainties for these data are caused by magnitude fluctuations in the secondary particle contamination and the uncertainty in $dN_{\rm ch}/d\eta$. These are followed by the systematic bias caused by multiplicity-dependent changes in the slope of the single-particle $p_T$ spectrum. Residual uncertainties from the pileup correction procedure and two-track inefficiency corrections contribute smaller systematic errors. Relative PV position event-mixing systematics are only significant in the upper $(y_{T1},y_{T2})$ corner of the acceptance for $y_T > 4$ where statistical errors dominate. The remaining systematic uncertainties discussed in this section were negligible, but were included.

All systematic offsets were summed linearly, while all systematic uncertainties were combined in quadrature and applied to the measured value plus offset, yielding asymmetric systematic uncertainty ranges in each $(y_{T1},y_{T2})$ bin. The uncertainty ranges in a few bins were extended to encompass the measured correlation value when necessary. The measured values in each $(y_{T1},y_{T2})$ bin were not corrected with the systematic offsets. 

In general, the total systematic uncertainties vary from about 10\% of the overall amplitude scale of the correlation structures in more-peripheral collisions to about 8\% in more-central. Systematic uncertainties exceed the statistical errors at lower transverse rapidity up to $y_T \approx 3$ or more; statistical errors dominate at larger $y_T$. The statistical errors and the systematic offset and uncertainties for the CI, all-azimuth mid-central 28-38\% correlations are shown in comparison with the correlations in Fig.~\ref{Fig4}. The corresponding statistical errors, and systematic offsets and uncertainties for the remaining data have similar structures and relative magnitudes as those shown in this figure and in more detail in Appendix~\ref{Appendix-D}.

\section{Theoretical Monte-Carlo Predictions}
\label{SecVI}

The predictions of two distinct theoretical approaches using {\sc hijing}~\cite{HIJING} and {\sc epos}~\cite{EPOS} were generated and compared to the charged-particle $p_T$ spectra and the CI, all-azimuth $(y_{T1},y_{T2})$ correlation data reported here. Both models include event-by-event dynamical fluctuations that generate correlations. The predicted $p_T$ spectra were fitted with Levy model distributions~\cite{Levy}. These parameters are listed in Table~\ref{TableI} and can be compared with the corresponding Levy model parameters used to fit the experimental STAR measurements that are listed in Table~\ref{TableII} (see Appendix~\ref{Appendix-A}). Comparisons of the predicted and measured correlations are discussed below. 

\begin{table*}[t]
	\caption{Centrality, average numbers of participant nucleons, NN binary collisions, and average multiplicity from Ref.~\cite{axialCI}. Centrality is also indicated with parameter $\nu = N_{\rm part}/(N_{\rm bin}/2)$~\cite{axialCI}. Parameters for the Levy model distribution representations (see Eq.~(\ref{EqA2})) of 200 GeV Au+Au minimum-bias $p_T$ spectrum data are also listed as explained in the text.}
\label{TableII}
\begin{tabular}{ccccc|ccc}
\hline \hline
 \multicolumn{5}{c}{Centrality \& MC-Glauber} & \multicolumn{3}{c}{Charge distribution}  \\
        Cent.(\%) & $\nu$ & $N_{\rm part}$ & $N_{\rm bin}$ & $dN_{\rm ch}/d\eta$ & $A_{\rm ch}$ & $T_{\rm ch}$ (GeV) & $q_{\rm ch}$  \\
\hline
84-93 & 1.40 & 4.6   & 3.2  & 5.2  & 14.78 & 0.1537 & 10.54  \\
74-84 & 1.68 & 10.5  & 8.8  & 13.9 & 36.15 & 0.1634 & 10.90  \\
64-74 & 2.00 & 20.5  & 20.5 & 28.8 & 69.65 & 0.1720 & 11.33  \\
55-64 & 2.38 & 36.0  & 42.8 & 52.8 & 119.7 & 0.1802 & 11.87  \\
46-55 & 2.84 & 58.1  & 82.5 & 89.  & 190.4 & 0.1882 & 12.56  \\
38-46 & 3.33 & 86.4  & 144  & 139. & 283.3 & 0.1953 & 13.32  \\
28-38 & 3.87 & 124.6 & 241  & 209. & 408.7 & 0.2018 & 14.19  \\
18-28 & 4.46 & 176.8 & 394  & 307. & 578.0 & 0.2080 & 15.16  \\
9-18  & 5.08 & 244.4 & 621  & 440. & 801.9 & 0.2136 & 16.22  \\
5-9   & 5.54 & 304.1 & 842  & 564. & 1006. & 0.2174 & 17.01  \\
0-5   & 5.95 & 350.3 & 1042 & 671. & 1176. & 0.2205 & 17.73  \\\hline \hline
\end{tabular}
\end{table*}

\begin{figure*}[t]
\includegraphics[keepaspectratio,width=7.0in]{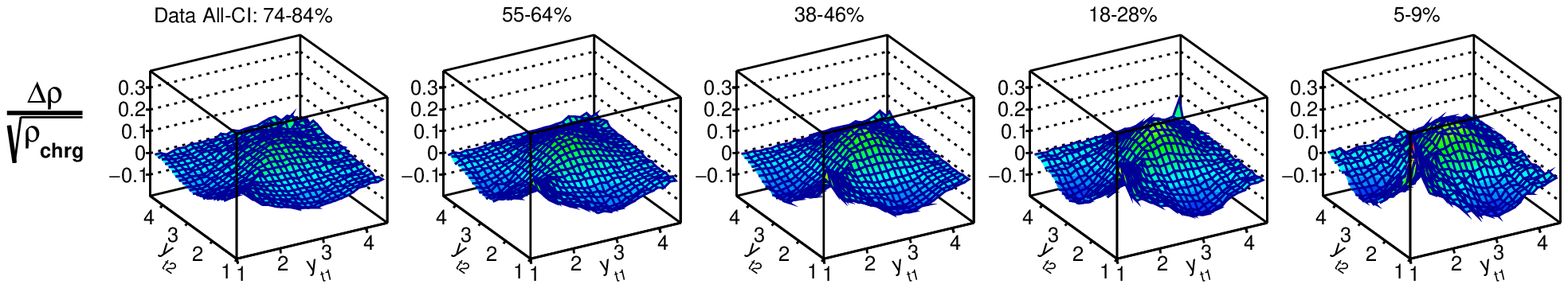}
\includegraphics[keepaspectratio,width=7.0in]{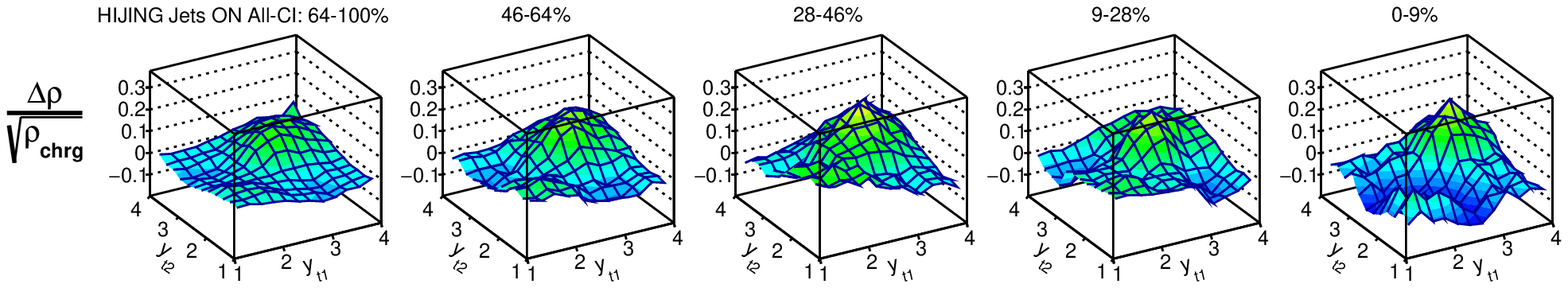}
\includegraphics[keepaspectratio,width=7.0in]{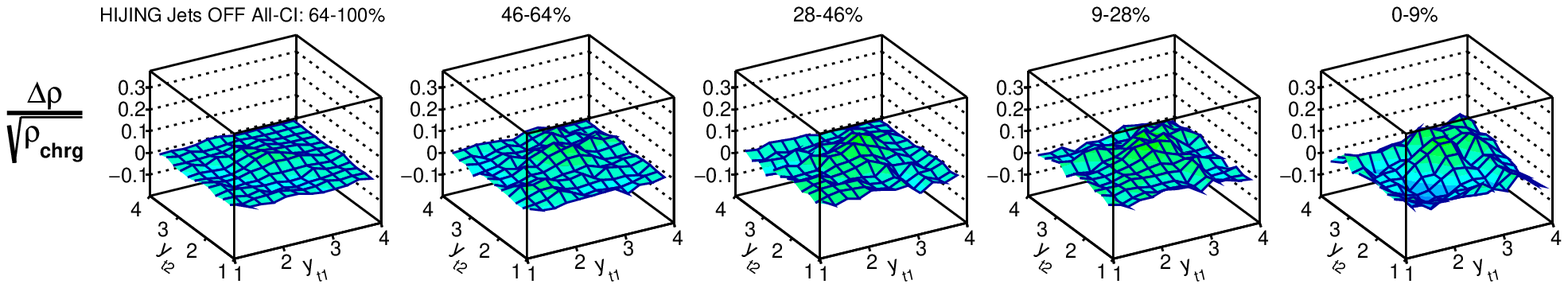}
\includegraphics[keepaspectratio,width=7.0in]{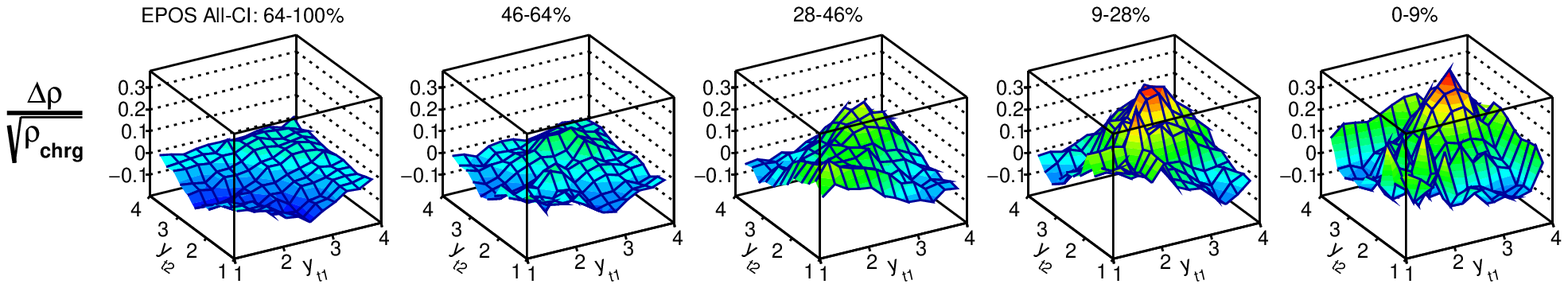}
\caption{\label{Fig5}
Comparisons between theoretical model predictions and measured two-dimensional correlations $\Delta\rho/\sqrt{\rho_{\rm chrg}}$ on coordinates $(y_{T1},y_{T2})$ for Au+Au collisions at $\sqrt{s_{\rm NN}}$ = 200 GeV for all charged particle pairs and all relative azimuthal angles. Data are shown in the upper row of panels for centrality cross section fractions 74-84\%, 55-64\%, 38-46\%, 18-28\% and 5-9\% from left-to-right, respectively. The next three rows show model predictions for 200~GeV Au+Au collisions with {\sc hijing} jets-on, {\sc hijing} jets-off, and {\sc epos}. Centralities for the {\sc hijing} and {\sc epos} predictions are shown in each row from left-to-right for the broader cross section fractions 64-100\%, 46-64\%, 28-46\%, 9-28\% and 0-9\%, respectively.}
\end{figure*}

\begin{figure*}[t]
\includegraphics[keepaspectratio,width=3.2in]{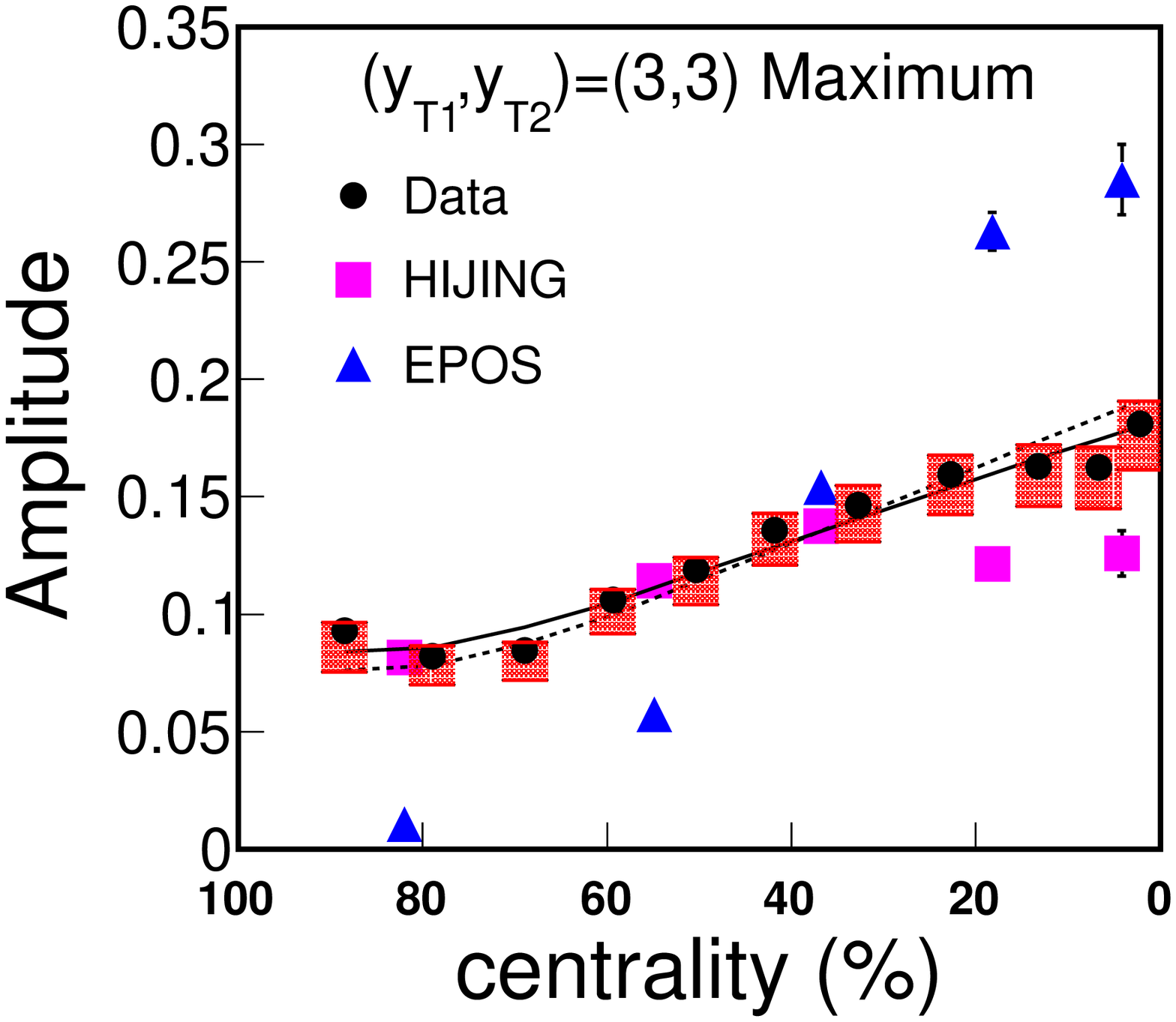}
\includegraphics[keepaspectratio,width=3.2in]{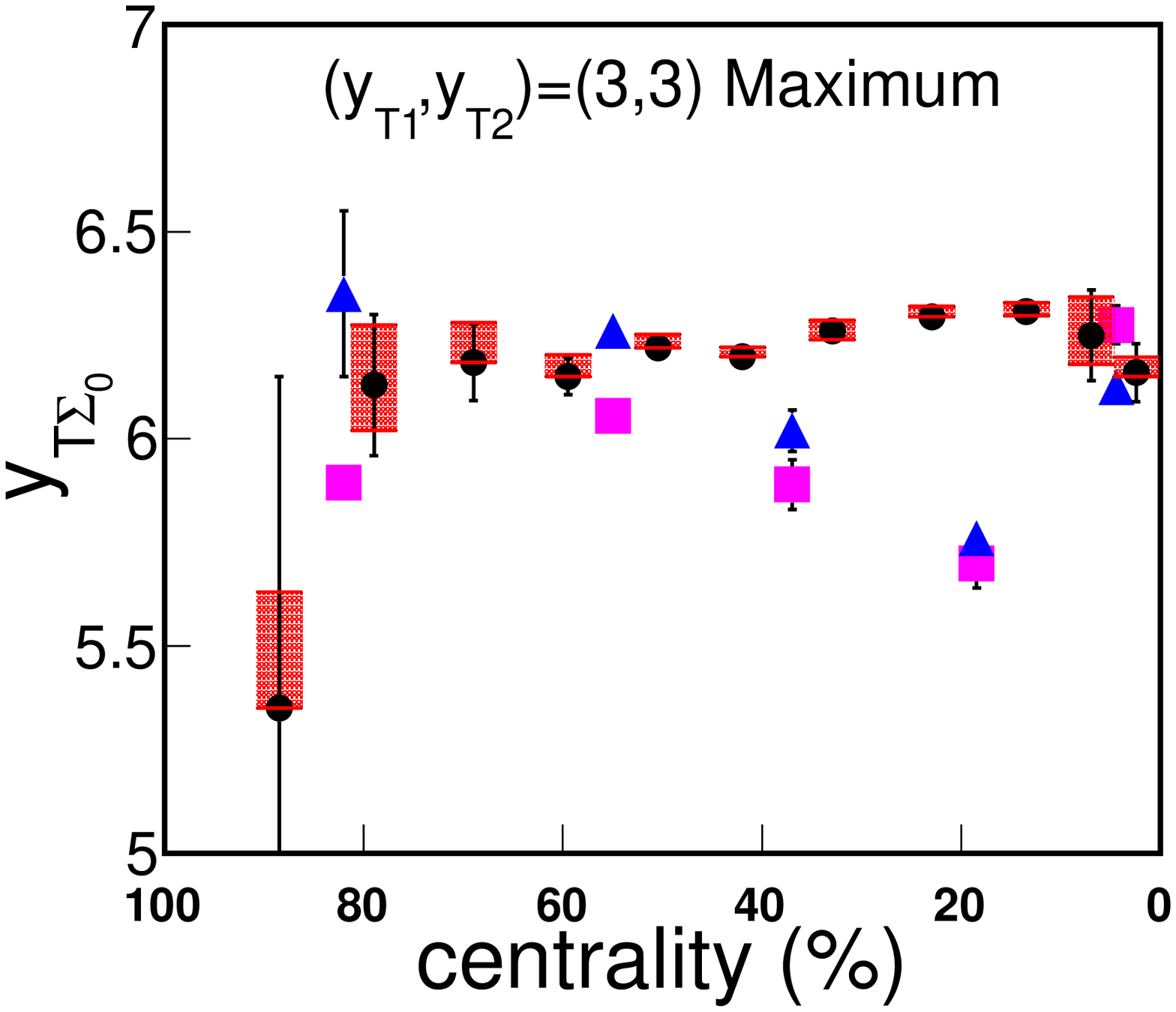}
\caption{\label{Fig6}
	Fit results for the amplitudes and positions of the measured and predicted correlation peak near $(y_{T1},y_{T2}) \approx$ (3,3) as a function of centrality. Centrality is denoted by total cross section percent and increases from peripheral to most-central from left to right. Peak position along the $y_{T1} = y_{T2}$ diagonal is denoted by the sum variable $y_{T\Sigma_0}$. Black, magenta, and blue data points indicate results for data, {\sc hijing} jets-on and {\sc epos}, respectively. Statistical errors are indicated by black error bars if larger than the symbols, while systematic uncertainties are shown as red shaded-boxes for the data. A binary scaling function fit (see text) to the measured correlation peak amplitudes is shown by the solid black curve. The fit requiring exact binary scaling ($\gamma = 1$) is shown by the dotted black curve.} 
\end{figure*}

\begin{figure*}[t]
\includegraphics[keepaspectratio,width=2.1in]{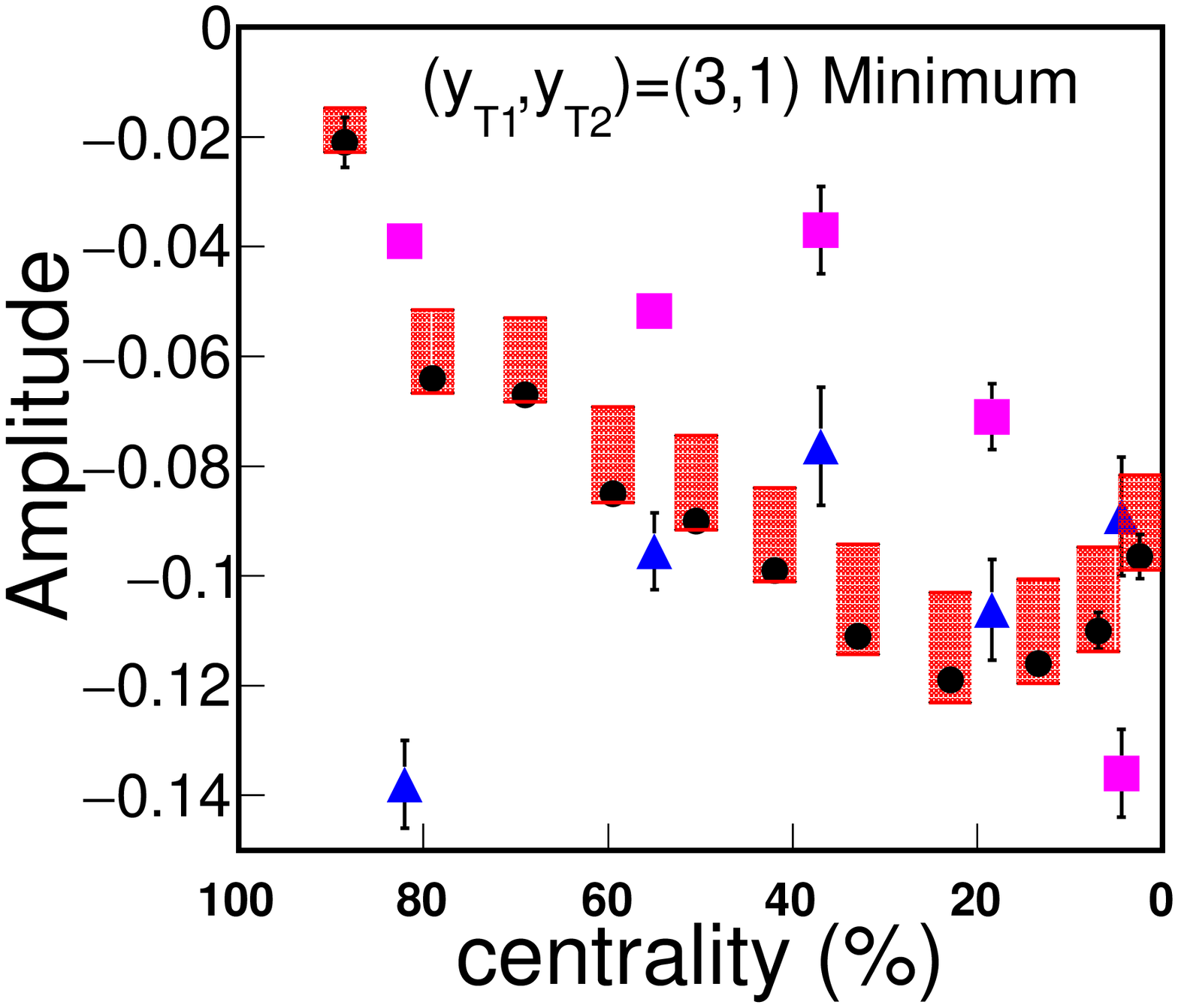}
\includegraphics[keepaspectratio,width=2.1in]{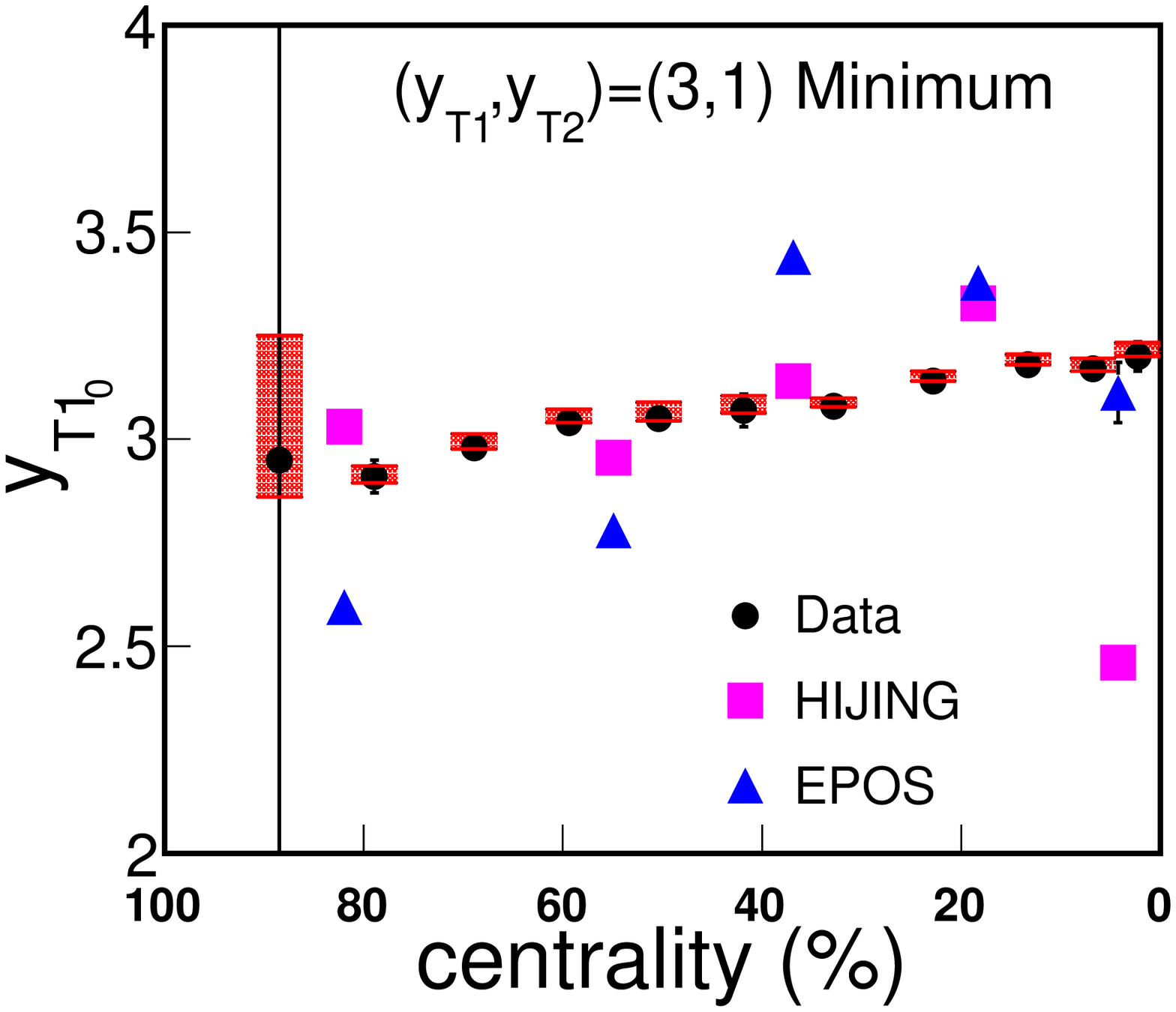}
\includegraphics[keepaspectratio,width=2.1in]{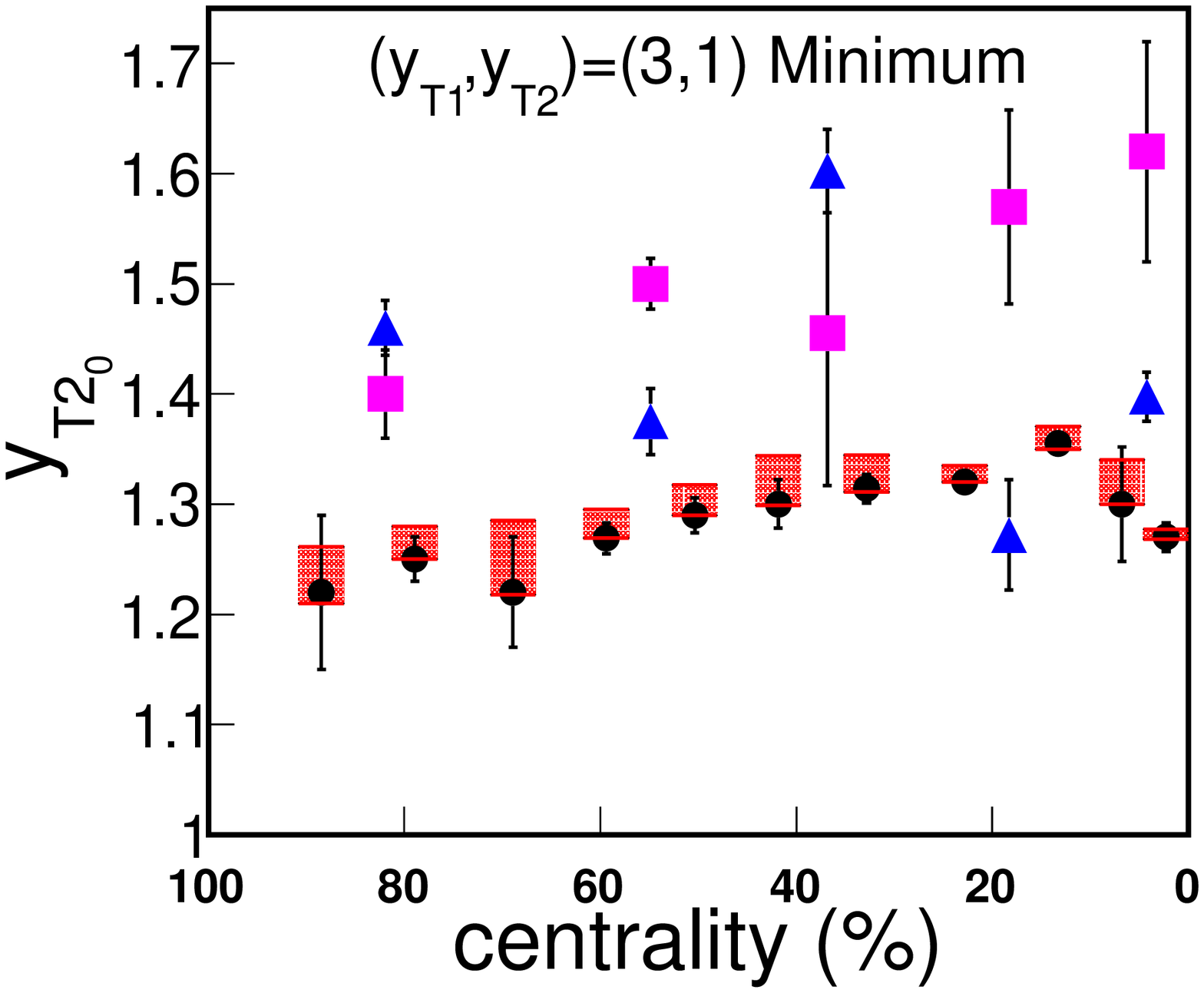}
\caption{\label{Fig7}
	Same as Fig.~\ref{Fig6} except for the amplitudes and positions of the saddle-shape minima. The positions are denoted by $y_{T1_0}$ and $y_{T2_0}$.}
\end{figure*}

\subsection{HIJING} 
\label{SecHIJING}

In {\sc hijing}, most particles are generated via color string + string soft collisions using the wounded nucleon~\cite{fritiof} and dual parton models (DPM)~\cite{DPM}, assuming binary collisions, and allowing both excitations and de-excitations of string masses. Color-strings are hadronized using the LUND~\cite{LUND} model. Fluctuating string mass leads to correlations that can increase with centrality due to multiple string + string collisions in {\sc hijing}. 

When semi-hard parton scattering and fragmentation (using PYTHIA~\cite{PYTHIA}) is included (jets on), fluctuations in the event-wise relative number of semi-hard produced particles generate a modest saddle-shaped correlation~\cite{phenom}. When the correlations among the particles in the jets increase, for example due to fluctuations in the number and/or energies of the jets, a saddle-shaped correlation is also produced but with an enhanced peak near $(y_{T1},y_{T2}) \approx (3,3)$ as shown in Ref.~\cite{phenom}. The latter correlation structure dominates the previous two weaker correlations. Particles produced in each fragmenting color-string and jet are combined independently in the final-state. {\sc hijing} provides a null hypothesis for particle production and correlations in heavy-ion collisions in the absence of an interacting medium.

For the present application two sets of minimum-bias Au+Au collision events at $\sqrt{s_{\rm NN}}$ = 200~GeV, using {\sc hijing} version 1.382, were generated where jets were either included or not included, referred to as ``jets on'' or ``jets off.'' Each set included 400K events.
The simulated events were binned into centrality selections based on charged-particle multiplicity within $|\eta| \leq 1$, full $2\pi$ azimuth, and $p_T \geq 0.15$~GeV/$c$, the same as was done for the data. The number of simulated {\sc hijing} collisions was chosen to be similar to the limited number of available {\sc epos} predictions described below. The resulting number of events was sufficient to achieve reasonable statistical accuracy for only five centrality bins, given by the total cross section ranges 0-9\%, 9-28\%, 28-46\%, 46-64\% and 64-100\% that were selected to overlap the centrality bins used for the data. Multiplicity-based centrality cuts were separately determined for both the jets-on (no quenching) and jets-off simulations. The 2D correlations were computed in the range $1.0 \leq y_T \leq 4.0$ and binned in a $12 \times 12$ uniform grid.

The Levy model parameters (see Appendix~\ref{Appendix-A}) that fit the {\sc hijing} jets-on and jets-off $p_T$ spectra are listed in Table~\ref{TableI}. The predicted charged-particle spectra, which are overall dominated by pions (88\%), were fit from $p_T$ = 0.15~GeV/$c$ to 4~GeV/$c$ (2.8 to 3.6 GeV/$c$ for jets-off). The jets-on parameters may be compared with those for measured spectra in Table~\ref{TableII} for similar centralities. The resulting Levy temperature and exponent for jets-on is similar to that for the measured spectra near mid-centrality, but does not predict the centrality dependence found for the measured spectra. With jets-off, the Levy temperatures are much too low and the Levy exponents much too large (tending to a Maxwell-Boltzmann distribution).

\begin{table*}[t]
\caption{\label{TableI}
Fit parameters for Levy model descriptions of {\sc hijing} and {\sc epos} predicted single-particle $p_T$ spectra at the $\chi^2$-minima. Listed in the columns are the centrality ranges (fraction of total cross section), number of participant nucleons (interpolated from Ref.\cite{axialCI}), charged particle multiplicities at mid-rapidity for $p_T > 0.15$~GeV/$c$, and Levy distribution fit parameters defined in Appendix~A, Eq.~(\ref{EqA2}) for the charged particle distributions $d^2N_{\rm ch}/dy_T d\eta$ for 200 GeV minimum-bias Au-Au collisions predicted by {\sc hijing} with jets-on, {\sc hijing} with jets-off, and {\sc epos}.}
\begin{tabular}{cc|cccc|cccc|cccc} \hline \hline
Cent. &  $N_{\rm part}$  & \multicolumn{4}{c}{{\sc hijing} jets-on} & \multicolumn{4}{c}{{\sc hijing} jets-off} & \multicolumn{4}{c}{\sc epos} \\
  &  & $dN_{\rm ch}/d\eta$ & $A_{\rm ch}$ & $T_{\rm ch}$ & $q_{\rm ch}$ & $dN_{\rm ch}/d\eta$ & $A_{\rm ch}$ & $T_{\rm ch}$ & $q_{\rm ch}$ & $dN_{\rm ch}/d\eta$ & $A_{\rm ch}$ & $T_{\rm ch}$ & $q_{\rm ch}$ \\
 (\%) & & & [(GeV/$c$)$^{-2}$] & (GeV) &  &  & [(GeV/$c$)$^{-2}$] & (GeV) &  &  & [(GeV/$c$)$^{-2}$] & (GeV) &  \\
\hline
64-100 &  10.72 & 10.5 & 30.6 & 0.171 & 12.6  & 6.36 & 30.5 & 0.144  & 66.1 & 18.5 & 44.3 & 0.195 & 15.4 \\
46-64  &  48.67  & 55.2 & 151  & 0.177 & 12.6  & 28.2 & 125  & 0.148  & 41.0 & 85.3 & 153  & 0.236 & 22.8 \\
28-46  &  108.4 & 139  & 364  & 0.181 & 12.36 & 62.4 & 258  & 0.1514 & 28.3 & 195  & 307  & 0.258 & 31.8 \\
9-28   &  212.2 & 301  & 757  & 0.185 & 12.4  & 123  & 481  & 0.1549 & 24.5 & 373  & 544  & 0.273 & 43.4 \\
0-9    &  330.4 & 517  & 1268 & 0.188 & 12.5  & 195  & 729  & 0.158  & 23.3 & 593  & 846  & 0.277 & 45.4 \\
\hline \hline
\end{tabular}
\end{table*}

The $(y_{T1},y_{T2})$ correlations were calculated as discussed in Sec.~\ref{SecII} for non-identified charged particles. The same-event and mixed-event pair distributions were calculated using the event averages
\bea
\rho_{{\rm se,HIJ},kl} & = & \frac{1}{\epsilon} \sum_{j=1}^{\epsilon} \frac{\bar{N}}{n_j}
n_{j,kl}^{\rm se}
\label{Eq82} \\
\rho_{{\rm me,HIJ},kl} & = & \frac{\bar{N}-1}{\bar{N}\epsilon_{\rm mix}} \sum_{j \neq j^{\prime}}
n_{jk} n_{j^{\prime}l}
\label{Eq83}
\eea
where the sums include all charged-particle pairs and all relative azimuthal angles. In these definitions, $\epsilon$ is the number of simulated events in the centrality bin, $\bar{N}$ is the mean charged-particle multiplicity in the acceptance, $n_j$ is the event-wise multiplicity, $n_{j,kl}^{\rm se}$ is the event-wise number of charged-particle pairs in event $j$ in $(y_{T1},y_{T2})$ bin $(k,l)$, $\epsilon_{\rm mix}$ is the number of simulated mixed-events, and $n_{jk}$ is the event-wise number of particles in arbitrary $y_T$ bin $k$. Equations~\ref{Eq82} and \ref{Eq83} were used for both the jets-on and jets-off correlations.

The final correlation quantity for either the jets-on or jets-off simulations is given by
\bea
\left( \frac{\Delta\rho}{\sqrt{\rho_{\rm chrg}}} \right)^{\rm HIJ}_{kl} & = &
{\cal P}^{\rm HIJ-CI,All}_{kl}
\left( \frac{\Delta\rho}{\rho_{\rm me}} \right)^{\rm HIJ}_{kl}
\nonumber \\
 &  & \hspace{-0.5in} = {\cal P}^{\rm HIJ-CI,All}_{kl}
\frac{ \rho_{{\rm se,HIJ},kl} - \rho_{{\rm me,HIJ},kl} }{ \rho_{{\rm me,HIJ},kl} }
\label{Eq84}
\eea
where the prefactors for the correlations were calculated with the corresponding predicted charged particle spectra within each centrality bin for either the jets-on or jets-off calculations. 


The event averages in Eqs.~(\ref{Eq82}) and (\ref{Eq83}) are susceptible to bias effects~\cite{MCBias} caused by event-mixing within multiplicity sub-bins that are too broad, the same as in the data analysis (see Sec.~\ref{SecV}). Due to the limited sample size in these simulations the multiplicity sub-bins could not be sufficiently reduced in width to completely eliminate this bias in the two most-central bins. The bias produced a constant offset~\cite{MCBias} in the quantity $[\Delta\rho/\rho_{\rm me}]^{\rm HIJ}_{kl}$, which for jets-on (jets-off) equaled 0.0014 and 0.005 (0.0016 and 0.004) for the 9-18\% and 0-9\% centrality bins, respectively. This bias offset was subtracted prior to multiplication by the prefactor in Eq.~(\ref{Eq84}).

The CI, all-azimuth predictions are shown and compared with data in Fig.~\ref{Fig5}. The data are shown in the upper row of panels for centralities 74-84\%, 55-64\%, 38-46\%, 18-28\% and 5-9\% from left-to-right. The second and third rows of panels show the {\sc hijing} predictions with jets turned on and off, respectively.

The overall saddle-shape and $(y_{T1}, y_{T2}) \approx (3,3)$ peak structures are apparent in the {\sc hijing} predictions. However, the {\sc hijing} model without jet production is completely inadequate for describing these correlations. The major features of the data and theoretical predictions are directly compared in Figs.~\ref{Fig6} and \ref{Fig7} which show the amplitudes and positions of the peak structure and the off-diagonal minima in the saddle structure, respectively. In Fig.~\ref{Fig6} the {\sc hijing} jets-on predicted amplitudes of the $(y_{T1}, y_{T2}) \approx (3,3)$ peak from most-peripheral to mid-central agree with the data, but fall below the measurements for more-central collisions. The predicted peak positions are similar to, but generally a few percent smaller than the data, except for the most-central collisions for which the predictions agree with the data. The predicted depths of the off-diagonal minima, shown in Fig.~\ref{Fig7}, are about one-half that of the data for most centralities, except the most-central bin. The predicted positions of the minima approximately agree with the data, however the prediction for the most-central collisions is considerably smaller than the data. In the most-peripheral 64-100\% bin the {\sc hijing} predicted correlations with jets-on are similar in shape and overall amplitude to the data. This may indicate that the event-wise fluctuation mechanisms in {\sc hijing}, longitudinal string and transverse parton fragmentations, are realistic, at least with respect to this type of correlation in peripheral collisions. Clearly, {\sc hijing} fails as the collision dynamics become more complicated with increasing centrality. 

\subsection{EPOS} 
\label{SecEPOS}

{\sc epos} version 3.210(c)~\cite{EPOS,KWGS} was used to provide hydrodynamic predictions for the CI, all-azimuth correlations. In this model the fluctuating initial collision stage is described in a multiple scattering framework using soft and hard Pomerons, including gluon saturation effects. Initial-stage interactions are separated into ``core'' and ``corona'' domains based on the transverse momentum and local density in the transverse plane~\cite{KlausWernerPRC} of the color-strings, or flux tubes, formed in the initial interactions. Subsequent evolution of the initial-stage core region, assumed to be a quark-gluon plasma (QGP), is described using (3+1)D viscous hydrodynamics until the hadronization stage. Final-stage, hadronic re-scattering and reactions for hadrons produced in both the core and corona regions are described using UrQMD~\cite{UrQMD}.

A total of 200K minimum-bias 200~GeV Au+Au collisions were generated~\cite{Anders} and separated into the five centrality bins used for the above {\sc hijing} predictions using the {\sc epos} predicted charged-particle multiplicities. Correlations were calculated using Eqs.~(\ref{Eq82}) and (\ref{Eq83}) above, where the predicted charge-particle spectra were used to calculate the prefactor. Statistical limitations restricted the multiplicity sub-binning, resulting in bias offsets of 0.001 and 0.0055 in the 9-28\% and 0-9\% centrality bins, respectively. The bias offsets were subtracted before multiplying by the prefactor. The {\sc epos} correlations were also binned on a $12 \times 12$ uniform grid from $y_T$ = 1.0 to 4.0.

The Levy model parameters that fit the {\sc epos} $p_T$ spectra are listed in Table~\ref{TableI}. The predicted charged-particle spectra, which are overall dominated by pions (85\%), were fit from $p_T$ = 0.15~GeV/$c$ to 4~GeV/$c$. The resulting Levy temperatures and exponents are generally higher than those describing the measured spectra, but both display monotonic increases with centrality as seen in the Levy parameter fits to the measured spectra.

Except for the most-peripheral collisions, the {\sc epos} predictions display the overall saddle-shape plus $(y_{T1}, y_{T2}) \approx (3,3)$ peak structure of the data. The predicted amplitude, shown in Fig.~\ref{Fig6}, is too low in more-peripheral collisions and increases too rapidly with centrality, being almost twice that of the data in the two more-central bins. The predicted peak positions approximately agree with the data and follow a similar trend on centrality as the {\sc hijing} jets-on predictions. The predicted amplitudes and positions of the saddle-shape minima, shown in Fig.~\ref{Fig7}, approximately agree with the data except for the most-peripheral collisions.

The {\sc epos} predictions for the most-peripheral collisions differ significantly from the data, indicating that for these collisions dynamical fluctuations in the assumed core and corona regions poorly represent those occurring in the collisions, at least as they affect the $(y_{T1},y_{T2})$ correlations. For the more-central collisions the hydrodynamic medium and/or the un-dissipated scatterings in the corona region are capable of generating realistic CI correlations on transverse rapidity. It would be beneficial to conduct additional correlation studies with {\sc epos} in which the relative sizes of the core and corona regions are varied, the hard-scattering processes are turned on and off, and final-stage hadronic rescattering is, or is not, included.

\section{Discussion}
\label{SecVII}
\subsection{Angular versus $(y_{T1},y_{T2})$ correlations}

Particle-number correlation-distributions on transverse rapidity are sensitive to the correlated fluctuations among the final-state particles and measure the co-variation between the numbers of particles at different transverse rapidities. On the other hand, particle number correlation distributions on relative angle, $\Delta\eta$ and/or $\Delta\phi$, are determined by the average number of correlated pairs produced by randomly distributed processes in the primary $(\eta,\phi)$ space. For example, elliptic flow occurs with respect to a randomly oriented event-plane, resulting in a $\cos{(2\Delta\phi)}$ correlation distribution. Randomly oriented dijets produce a NS 2D peak and an AS ridge distribution on $\Delta\eta,\Delta\phi$ that contain information about the event-average number of correlated pairs for all the dijets in a collision. However, if the same number of randomly distributed dijets occurs in each event, each of which has the same energy and fragment distribution, then the resulting pair-number correlations on transverse rapidity will be zero due to the absence of fluctuations. In other words, the same-event and mixed-event pair distributions would be the same.

\subsection{Sources of $(y_{T1},y_{T2})$ correlations}

Dynamical processes that affect the event-wise single-particle, transverse rapidity parent distribution will generate correlation structure on $(y_{T1},y_{T2})$. For example, at fixed multiplicity, event-wise fluctuations in the overall slope of the parent $p_T$-spectrum, $dN_{\rm ch}/p_T dp_T$, cause the resulting set of parent distributions to pivot about an intermediate $p_T$ like a see-saw. The number of sampled particle-pairs in either lower $p_T$ bins or in higher $p_T$ bins increase or decrease together, relative to the mean, resulting in positive covariance. Conversely, pairs with one particle in a lower $p_T$ bin and the other in a higher $p_T$ bin have negative covariance. The result of such processes is a saddle-shaped correlation in 2D space. In {\sc hijing}, this can occur when the number and/or rest energies of the color-strings fluctuate within each collision and/or from event-to-event. In {\sc epos}, similar color-string fluctuations in the corona region plus dynamical fluctuations in the freeze-out temperature from the core region can produce similar saddle-shaped correlation distributions.

Other dynamical processes may be more effective in specific regions of $p_T$. Fluctuations in transverse flow from varying initial conditions affect the curvature of the $p_T$ spectrum at higher $p_T$ and also produce a saddle-shaped correlation, but with a different shape than that from fluctuations in overall slope. Fluctuations in the number, energies and fragment distributions of jets affect the $p_T$ spectrum at higher $p_T$ and also produce a saddle-shaped correlation, but one having a distinctive enhancement at intermediate and higher $p_T$. Examples of the saddle-shaped correlations from these sources are shown in Ref.~\cite{phenom} based on a phenomenological model. However, the saddle-shape is not expected to continue increasing at higher $p_T$ even if all particles at higher momentum are correlated, e.g., from the same hard jet, thermal hot-spot, or high velocity outgoing plume of particles from a localized, initial high pressure region. In such cases the ratio of the number of correlated pairs to mean multiplicity, $\Delta\rho/\sqrt{\rho_{\rm chrg}}$ is proportional to $dN_{\rm ch}/dy_T$, and therefore falls off with $y_T$. The expected correlations from the dynamical processes included in {\sc hijing} and {\sc epos} are saddle-shapes that reach a peak at some intermediate $y_T$, then fall off at higher $y_T$.  

\subsection{Correlation distribution morphology}

The structures and centrality trends in the correlation measurements shown in Figs.~\ref{Fig1} - \ref{Fig3} merit further discussion. As shown in Fig.~\ref{Fig5} the CI, all-azimuth $(y_{T1}, y_{T2}) \approx (3,3)$ correlation peak (see Fig.~\ref{Fig1}) can be produced both in {\sc hijing} with jets-on and in {\sc epos} in most-central collisions where the ratio of core effects (hydro) to corona effects (jets) are maximum. The corresponding saddle-shape and its minima are also produced in both theoretical models.

The detailed correlation structures and shapes shown in Fig.~\ref{Fig2} are consistent with similar correlation measurements for minimum-bias p+p collisions at $\sqrt{s}$ = 200~GeV~\cite{Porter} when compared with the 84-93\% Au + Au correlations. The centrality evolution of each correlation structure in each charge-sign and azimuthal angle projection is smooth. The correlation peak at larger $y_T \approx 3$ in the NS-US correlations should be particularly sensitive to transverse fragmentation from jets or to other hadronization processes in heavy-ion collisions. Resonance decays will also contribute to this correlation projection. Resonance decay contributions are discussed in Appendix~\ref{Appendix-B} where they are shown to be about one-tenth the amplitude of the observed structures in the NS-US correlations for $y_T < 3$. The enhanced NS-US correlation peak amplitude near $y_T \sim 3$, relative to the NS-LS peak amplitude, resulting in the negative NS-CD correlation in Fig.~\ref{Fig3} is consistent with a transverse parton scattering and fragmentation mechanism~\cite{PYTHIA} that produces charge-ordering~\cite{LUND,ISR}. These NS-CD correlations provide new constraints on fragmentation and recombination models of hadronization~\cite{RudyHwa}.

In addition, the second maximum at $y_T \approx$ 2.0 to 2.5 in the NS-US correlations is intriguing. The corresponding structure in the NS-LS correlations, if present, is obscured by the HBT correlations. At any rate, the US structure is larger than the LS, if the latter exists at all. Charge-ordering among soft particle production from longitudinal color-strings (LUND) or from charge-ordered hadronization of a Bjorken-expanding, hydrodynamic medium~\cite{Bjorken} might account for these structures. However, corresponding angular correlations~\cite{axialCI} for these same data show an expected structure from such longitudinal, charge-ordered processes that quickly dissipates within the first few peripheral collision bins. The results in Fig.~\ref{Fig2} for NS-US show the second peak at $y_T \approx$ 2.0 to 2.5 begins to appear in the 64-74\% bin and then steadily increases to most-central collisions. It is noteable that the ridge correlation in angular correlations for these same data follow a similar centrality trend, where there is
some indication that US ridge amplitudes exceed the LS amplitudes~\cite{SamRonPaper,LizThesis}. Determining the relation, if any, between the ridge correlation observed in angular correlations and this peak at $y_T \approx$ 2.0 to 2.5 in the NS-US correlations requires further analysis beyond the present scope of this study.

The AS-LS and AS-US correlation peaks at (3,3) and their shapes, either symmetric (for LS) or asymmetric (for US), require detailed consideration of initial-state partonic $k_T$, jet-quenching and jet-broadening on these back-to-back correlations~\cite{Hirano,Mueller}. The corresponding hydrodynamic effects on back-to-back correlations, e.g. medium recoil and diffusion wakes, would be interesting to compare with these data. The subsidence and re-emergence of the AS-US correlation peak at low $y_T \approx 1$, in going from most-peripheral to most-central collisions, is unusual. The reduction with increasing centrality might be expected for longitudinal fragmentation/hadronization as discussed above, but the increase for more-central collisions is not understood.  

The negative, NS-CD correlations shown in Fig.~\ref{Fig3}, are likely a result of a hadronization mechanism, such as fragmentation, that produces more US pairs at nearby relative angles and relative transverse rapidity $y_{T\Delta}$ than LS pairs~\cite{LUND,ISR}. The positive AS-CD correlations at lower $y_T \approx 2.2$ reflect the suppression in the AS-US correlations in this lower $y_T$ range shown in Fig.~\ref{Fig2}. The varied structures and centrality evolution shown in these charge- and $\Delta\phi$ projection-dependent correlations, together with the corresponding angular correlations as a function of the transverse momenta of the two charged-particles, provide significant new constraints on models of relativistic heavy-ion collisions.

The amplitudes of the measured and predicted correlation maxima near $(y_{T1},y_{T2}) \approx (3,3)$ and the saddle-shape minima near $(y_{T1},y_{T2}) \approx (3,1)$ plus the positions of the maxima and minima
were determined using simple model fits (2D second-order polynomial or 2D Gaussian) to several bins located near those bins having the local maximum or minimum correlation amplitude. The results are presented in Figs.~\ref{Fig6} and \ref{Fig7}. The peak amplitudes in the measured correlations are well described with a simple, binary scaling function given by $A [N_{\rm bin}/(dN_{\rm ch}/d\eta)]^{\gamma}$ using the values of $N_{\rm bin}$ and $dN_{\rm ch}/d\eta$ in Table~\ref{TableII}. The fitted functions are shown in Fig.~\ref{Fig6} by the dotted curve with $\gamma = 1$ (fixed), $A = 0.123 \pm 0.003$, and p-value = 0.66, and by the solid curve with fitted $\gamma = 0.83 \pm 0.09$, $A = 0.125 \pm 0.003$, and p-value = 0.91. 

\subsection{Theoretical predictions}

Comparison of the {\sc hijing} predictions without jets and with jets (but no jet quenching) to each other and to the data in Fig.~\ref{Fig5} clearly shows that within this purely scattering and fragmentation approach, jets are essential for describing the observed $(y_{T1},y_{T2})$ correlations. It is interesting to note from Fig.~\ref{Fig6} that the measured peak amplitudes near $(y_{T1},y_{T2}) \approx (3,3)$ follow binary NN collision scaling over the entire centrality range while the {\sc hijing} jets-on predictions are consistent with binary scaling from most-peripheral to mid-central collisions only, then level off and fall below binary scaling. Jet quenching is not included in these {\sc hijing}, jets-on predictions. The observed peak amplitudes are consistent with the {\sc hijing} predictions until mid-centrality, then exceed the predictions. It is also worth noting that the minima in the saddle-shaped correlations for the data are generally deeper than those in the {\sc hijing} jets-on predictions. Both of these deficiencies in the {\sc hijing} jets-on predictions suggest the need to include additional correlation effects due to the increasingly dense medium produced in the collision.

In {\sc epos}, correlations arise from fluctuations in the initial-state energy/momentum spatial distribution that evolve via hydrodynamics to the final-state, from soft and semi-hard scattering and fragmentation in the corona region, and in the final hadron-scattering stage. The predicted centrality dependence of the $(y_{T1},y_{T2}) \approx (3,3)$ peak amplitude is too large compared to data. For the more-peripheral collisions {\sc epos} under-predicts the peak amplitudes in contrast to the {\sc hijing} jets-on predictions, suggesting that the relative corona-to-core region contributions are too small in this centrality range. For mid- to most-central collisions the {\sc epos} predicted peak amplitude is in fair agreement with the trend of the data, but overestimates the amplitude for more-central collisions, suggesting excessive temperature and/or transverse flow fluctuations from the hydrodynamic core. Except for the most-peripheral (64-100\%) centrality bin, {\sc epos} predicts the saddle-shape minima depth and location fairly well.

Taken together, the {\sc hijing} jets-on and {\sc epos} predictions for the CI, all azimuth pair-number correlations on $(y_{T1},y_{T2})$, in comparison with the measurements, suggest that for Au+Au collisions at 200~GeV the present peripheral to mid-central collision correlations can be described, to first-order, as a minimally-interacting, superposition of NN collisions. For mid- to most-central collision systems the poor quality of the {\sc hijing} jets-on predictions suggests the presence of significant medium effects. The {\sc epos} model, which includes a strongly interacting core, predicts larger correlation magnitudes in this mid- to most-central range. This over-arching view is consistent with previous observations of non-identified, two-particle jet-like angular correlations for this same collision system and energy~\cite{axialCI}.

\section{Summary and conclusions}
\label{SecVIII}

Measurements of non-identified, charged-particle pair-number 2D correlation projections onto transverse rapidity were presented for minimum-bias Au+Au collisions at $\sqrt{s_{\rm NN}}$ = 200~GeV from the STAR Collaboration. Correlations were constructed for each of the four charge-pair combinations (LS, US, CI, and CD), for three relative azimuthal angle projections, and for eleven centrality bins. The overall correlation structure displays a saddle-shape and an enhanced, positive correlation peak near $(y_{T1},y_{T2}) \approx$ (3,3). Both features are expected. In addition, the measurements display considerable structure that varies significantly between LS and US charge combinations and for different relative azimuthal angle projections. Each of the correlation structures observed in these data evolves smoothly with centrality. The present measurements and analysis provide access to complementary information about the relativistic heavy-ion collision system compared to that which can be studied with angular correlations, the latter being sensitive to, for example, the per-event, average number of jet-like particle pairs or the average number of collectively flowing pairs. As such the present correlations enable novel tests of theoretical models.

The CI, all relative azimuthal angle correlation data were compared with theoretical predictions. Comparisons with {\sc hijing}, both with and without jets, and with {\sc epos}, an event-by-event (3+1)D hydrodynamic model, were presented in both visual and quantitative formats. 

{\sc hijing}, with longitudinal color-string fragmentation only (jets-off), does not generate correlations with sufficient amplitude. {\sc hijing} with jets-on predicts the major correlation structures in the data but with varying success with respect to the amplitudes. The amplitudes of the correlation peak near $(y_{T1},y_{T2}) \approx$ (3,3) for the peripheral to mid-central collisions are correctly predicted. However, {\sc hijing} with jets-on and no jet-quenching fails to achieve the larger amplitudes of the correlation peak in the more-central bins. In addition, the {\sc hijing} predictions reproduce only about one-half of the observed amplitudes of the saddle-shaped correlation structures that, in this model, are affected by correlated fluctuations in the color-string interaction and fragmentation process. Both deficiencies imply, not unexpectedly, that additional or stronger interactions with an increasingly dense medium must be accounted for. 

{\sc epos} also predicts each of the dominant correlation structures and predicts large amplitudes for the $(y_{T1}, y_{T2}) \approx (3,3)$ peak in the mid- to most-central collisions relative to data. However, the {\sc epos} predictions for the $(y_{T1}, y_{T2}) \approx (3,3)$ peak amplitude in more-peripheral collisions fall well below the data. Additionally, the increase in the peak amplitude with centrality is much greater than seen in the data. The predicted amplitudes and positions of the saddle-shape minima are in fair agreement with the data except in the most-peripheral centrality bin. Further study of the origin of the $(y_{T1},y_{T2})$ correlations predicted by {\sc epos}, including the CD and NS versus AS structures, is warranted in sophisticated models including fluctuating initial-states, hydrodynamics and parton energy loss. 

The results presented here suggest that, to first-order, the CI, all relative azimuth-angle $(y_{T1},y_{T2})$ correlations for peripheral to mid-central Au+Au collisions at 200~GeV can be described as a superposition of NN soft plus semi-hard collisions with minimal effects from the medium. For mid- to most-central collision systems significant medium effects are indicated. This over-arching view is consistent with previous observations of non-identified, two-particle jet-like angular correlations for this same collision system and energy~\cite{axialCI}. The full set of $(y_{T1},y_{T2})$ correlations reported here can be used in future efforts to further constrain theoretical models and improve the understanding of the dense, partonic system created in relativistic heavy-ion collisions at RHIC.

\vspace{0.2in}
{\bf Acknowledgements}
\vspace{0.1in}

The authors would like to thank Prof. Klaus Werner and Dr. Gabriel Sophys for their assistance in providing the {\sc epos} simulations and for discussions regarding the {\sc epos} model. We also thank Prof. Thomas Trainor for laying the foundation for the correlation quantities used in this paper and to Dr. Duncan Prindle for the pileup and two-track inefficiency corrections used here.
We thank the RHIC Operations Group and RCF at BNL, the NERSC Center at LBNL, and the Open Science Grid consortium for providing resources and support.  This work was supported in part by the Office of Nuclear Physics within the U.S. DOE Office of Science, the U.S. National Science Foundation, the Ministry of Education and Science of the Russian Federation, National Natural Science Foundation of China, Chinese Academy of Science, the Ministry of Science and Technology of China and the Chinese Ministry of Education, the Higher Education Sprout Project by Ministry of Education at NCKU, the National Research Foundation of Korea, Czech Science Foundation and Ministry of Education, Youth and Sports of the Czech Republic, Hungarian National Research, Development and Innovation Office, New National Excellency Programme of the Hungarian Ministry of Human Capacities, Department of Atomic Energy and Department of Science and Technology of the Government of India, the National Science Centre of Poland, the Ministry  of Science, Education and Sports of the Republic of Croatia, RosAtom of Russia and German Bundesministerium f\"ur Bildung, Wissenschaft, Forschung and Technologie (BMBF), Helmholtz Association, Ministry of Education, Culture, Sports, Science, and Technology (MEXT) and Japan Society for the Promotion of Science (JSPS).

\appendix
\section{Correlation prefactor}
\label{Appendix-A}

The prefactor for the CI combination using all pair-wise, relative azimuthal angles from $-\pi$ to $\pi$ is given in binned transverse-rapidity space by
\bea
{\cal P}^{\rm CI,All}_{kl} & \equiv & 
\left[ \frac{d^2N_{\rm ch}}{dy_{T1} d\eta} \frac{d^2N_{\rm ch}}{dy_{T2} d\eta} \right]^{1/2}
\label{EqA1}
\eea
where $y_{T1}$ and $y_{T2}$ are set equal to the mid-points of bins $k$ and $l$, respectively (see Sec.~\ref{SecII}). In Eq.~\ref{Eq2a} in Sec.~\ref{SecII} the prefactor is defined as the geometric mean of the product of event-average particle numbers in bins $k$ and $l$, corresponding to two-times (2$\times$) the preceding form for the present 2 units of pseudorapidity acceptance. With the preceding definition the maximum amplitude range for the correlations is [$-$0.5,0.5] corresponding to fully anti-correlated and fully correlated limits, respectively. 

In the above equation the charged particle distribution was parametrized with a Levy distribution~\cite{Levy}  given by
\bea
\frac{d^2N_{\rm ch}}{dy_T d\eta} & = & 2\pi p_T \frac{dp_T}{dy_T} \left[
\frac{d^2N_{\rm ch}}{2\pi p_T dp_T d\eta} \right] \nonumber \\
 & = & \frac{2\pi p_T m_T A_{\rm ch}}{\left[1+(m_T - m_0)/(T_{\rm ch}q_{\rm ch}) \right]^{q_{\rm ch}}}.
\label{EqA2}
\eea
Charged-particle spectra data corresponding to the centrality definitions used here are not available. Instead, the published spectra data were interpolated to the present centralities using the following steps. Transverse momentum spectra data for 200~GeV Au+Au minimum-bias collisions from the STAR Collaboration~\cite{STARspectra} were fitted for each available centrality from the lowest measured $p_T$ value to about 5~GeV/$c$ using the above Levy distribution. The $T_{\rm ch}$ and $q_{\rm ch}$ parameter distributions as functions of centrality were separately fit with power-law functions and interpolated to the centrality bin mid-points used in this analysis. The amplitudes $A_{\rm ch}$ were determined by requiring the integrated yields from $p_T$ = 0 to $\infty$ to equal the efficiency- and background-corrected yields, $dN_{\rm ch}/d\eta$, at each centrality, given in Table~III of Ref.~\cite{axialCI}. The resulting parameters $A_{\rm ch}$, $T_{\rm ch}$ and $q_{\rm ch}$ for the present 200~GeV Au+Au analysis are listed in Table~\ref{TableII}.

For LS and US correlations, the prefactor is reduced by $1/\sqrt{2}$ because there are one-half as many particle pairs available compared to using all charged-particle pairs. When the relative azimuthal angular range is restricted to either NS pairs or AS pairs, the prefactor is also reduced by $1/\sqrt{2}$.  The appropriate number of factors of $1/\sqrt{2}$ are applied to each of the charged-pair and azimuthal-angle range selections required for the correlation data presented in this paper.


\section{Secondary particle correlation uncertainty}
\label{Appendix-C}

	Estimates of the secondary particle contamination contributions to the $(y_{T1},y_{T2})$ correlations are given in this appendix. Contributions arising from event-wise fluctuations in the relative secondary-to-primary particle yield ratio and from event-wise fluctuations in the shape of the secondary particle $p_T$ distribution are included.

The observed single-particle distribution for an arbitrary centrality bin was assumed to be given by
\bea
\frac{d^2N}{dy_Td\eta} & = & \bar{N} [ (1 - \bar{\kappa}) \hat{\rho}_{\rm prim}(y_T)
                         + \bar{\kappa} \hat{\rho}_{\rm sec}(y_T) ]
\label{EqB1}
\eea
for mean charged-particle multiplicity $\bar{N}$, where mean secondary particle fraction $\bar{\kappa} = 0.12$~\cite{PRC79}. Primary and secondary particle distributions (discussed below), normalized to unity within the acceptance, are denoted by $\hat{\rho}_{\rm prim}$ and $\hat{\rho}_{\rm sec}$, respectively. 

For fluctuating secondary particle yields, the same-event pair distribution, assuming fixed shapes for the primary and secondary particle spectra, is given by
\bea
\frac{d^4N_{12}}{dy_{T1}d\eta dy_{T2}d\eta} & = & \bar{N}(\bar{N} - 1) \nonumber \\
 & & \hspace{-0.8in} \times \left\{
    [(1 - \bar{\kappa})^2 + \sigma^2_{\kappa}] \hat{\rho}_{\rm prim}(y_{T1}) \hat{\rho}_{\rm prim}(y_{T2})
    \right. 
\nonumber \\
 & &  \left. \hspace{-0.8in} + (\bar{\kappa}^2 + \sigma^2_{\kappa}) \hat{\rho}_{\rm sec}(y_{T1}) \hat{\rho}_{\rm sec}(y_{T2}) \right.
\nonumber \\
 & &  \left. \hspace{-0.8in} + [\bar{\kappa}(1 - \bar{\kappa}) - \sigma^2_{\kappa})]
   \left( \hat{\rho}_{\rm prim}(y_{T1}) \hat{\rho}_{\rm sec}(y_{T2})
\right. \right. \nonumber \\
	& &  \left. \left. \hspace{-0.8in}      +  \hat{\rho}_{\rm prim}(y_{T2}) \hat{\rho}_{\rm sec}(y_{T1}) \right) \right\}
\label{EqB2}
\eea
where $\sigma^2_{\kappa}$ is the variance of event-wise, secondary particle fraction $\kappa$. Correlations are produced when there are event-wise fluctuations in $\kappa$ and $\hat{\rho}_{\rm prim}(y_T) \neq \hat{\rho}_{\rm sec}(y_T)$. The correlations produced by fluctuations in $\kappa$, for CI, all-azimuth correlations, are given by
\bea
	\frac{\Delta\rho}{\sqrt{\rho_{\rm chrg}}} (y_{T1},y_{T2})_{\rm sec} & = & {\cal P}_{12}^{\rm CI,All}(y_{T1},y_{T2}) 
\nonumber \\
 & & \hspace{-1.0in} \times \left[ \frac{\frac{d^4N_{12}}{dy_{T1}d\eta dy_{T2}d\eta}}
      {\frac{\bar{N} - 1}{\bar{N}} \frac{d^2N}{dy_{T1}d\eta} \frac{d^2N}{dy_{T2}d\eta} } - 1 \right].
\label{EqB3}
\eea
Binned distributions are estimated by evaluating the continuous distributions at $y_T$-bin mid-points.

Ref.~\cite{PRC79} gives the pion and proton contamination fractions as functions of $p_T$ for the STAR Run~4 Au+Au 62.4~GeV collision data. We assumed the same contamination fractions for the 200~GeV data because the same detector configuration was used at both 62.4~GeV and 200~GeV during Run~4. The secondary kaon and anti-proton contamination fractions were negligible.
The secondary particle distribution was assumed to be
\bea
\frac{d^2N_{\rm sec}}{dy_Td\eta} & = & 2\pi p_T m_T \frac{d^2N{\rm ch}}{2\pi p_T dp_T d\eta}
\nonumber \\
 & \times & \left[ f_\pi F_{\rm sec}^{\pi^{\pm}}(p_T) + f_p F_{\rm sec}^{\rm proton}(p_T) \right]
\label{EqB6}
\eea
where $f_\pi$ = 0.85, $f_p$ = 0.033, $dp_T/dy_T = m_T$ at mid-rapidity, and the charged-particle distribution was represented with the Levy distribution in Appendix~\ref{Appendix-A} using the parameters in Table~\ref{TableII}. The pion and proton fractional background distributions were parametrized as,
\bea
F_{\rm sec}^{\pi^{\pm}}(p_T) & = & 0.04 + 0.155 e^{-3.57(p_T - 0.15)} \\
\label{EqB4}
F_{\rm sec}^{\rm proton}(p_T) & = & -1.15(p_T - 0.15) + 0.65, \nonumber \\
  & & {\rm for} \hspace{0.1in} 0.15\leq p_T \leq 0.575
\\
\label{EqB5}
 & = & 0.153 e^{-8(p_T - 0.575)}, ~~ p_T > 0.575
\label{EqB7}
\eea
for $p_T$ in units of GeV/$c$. Normalizing $d^2N_{\rm sec}/dy_Td\eta$ over the $y_T$ range from 1.0 to 4.5 to unity gives $\hat{\rho}_{\rm sec}(y_T)$. 
	Poisson fluctuations of ratio $\kappa$ were assumed for fixed, total multiplicity $\bar{N}$, where $\kappa = N_{\rm sec}/\bar{N}$, and $\sigma^2_{\kappa} = (\Delta N_{\rm sec})^2/\bar{N}^2$ $ \approx N_{\rm sec}/\bar{N}^2$ $ = \bar{\kappa}/\bar{N}$. From Table~\ref{TableII}, the Poisson fluctuations in secondary particle yields relative to $\bar{N}$ varies from 11\% in peripheral collisions to 1\% in most-central collisions.

	The estimated secondary-particle contamination magnitude was treated as an uncertainty, rather than a correction. One-half of $\Delta\rho / \sqrt{\rho_{\rm chrg}} (y_{T1},y_{T2})_{\rm sec}$ in each $(k,l)$ bin was assumed a systematic offset, and $\pm 1/2$ was assumed for the systematic uncertainty following the same procedure described in Sec.~\ref{SecV}. Because the secondary particle contamination fraction ($\kappa$) does not change significantly with centrality~\cite{PRC79}, these systematic uncertainties in the final correlations are also approximately constant with centrality.

Correlations for secondary particles also occur when the shapes (e.g., overall slopes on $p_T$) of the $\hat{\rho}_{\rm sec}(y_T)$ distributions fluctuate (see Sec.~\ref{SecVII}). Secondary particles, being predominately from weak-decays and from particle production in the detector material, would be expected to maintain, to some extent, the momentum correlations of their parent particles, which are primary particles from the collision. However, due to the weak-decay Q-values for $K^0_s \rightarrow \pi^+ \pi^-$ and $\Lambda \rightarrow p\pi$, and to the momentum transfers involved in secondary particle production processes in the detector material, we expect the correlations involving secondary particles to be diminished in amplitude and dispersed in relative angle compared to that for primary particle pairs.

Because the secondary particle contamination is dominant at lower $p_T$~\cite{PRC79}, we assumed the analytical 2D-Levy model, derived for the lower momentum range $p_T \leq 2$~GeV/$c$ in Ref.~\cite{Ayamtmt}, for both the primary and secondary same-event particle pair distributions. The 2D-Levy model was used in Ref.~\cite{Ayamtmt} to describe the transverse momentum correlations for 130~GeV Au+Au collisions. The parameters of the 2D-Levy model for preliminary, 200~GeV Au+Au $(y_{T1},y_{T2})$ correlations were taken from Ref.~\cite{LizThesis}. To account for the expected reduction in secondary particle correlation amplitudes, a 30\% reduction was assumed for the relative variance difference parameters $\Delta (1/n)_{\Sigma}$ and $\Delta (1/n)_{\Delta}$ defined in Ref.~\cite{Ayamtmt}. The relative variance differences determine the curvatures of the saddle-shaped correlation structure at the origin. The reduction in secondary particle correlations was based on the relative magnitudes of the above Q-values and the mean-$p_T$ for light-flavor particle production. This 30\% reduction produced very small effects on the final correlations, being about one-tenth of that produced by the above fluctuations in secondary particle yields. Although very small, these systematic uncertainties were included in quadrature in the total systematic uncertainty estimates.


\begin{figure*}[t]
\includegraphics[keepaspectratio,width=6.0in]{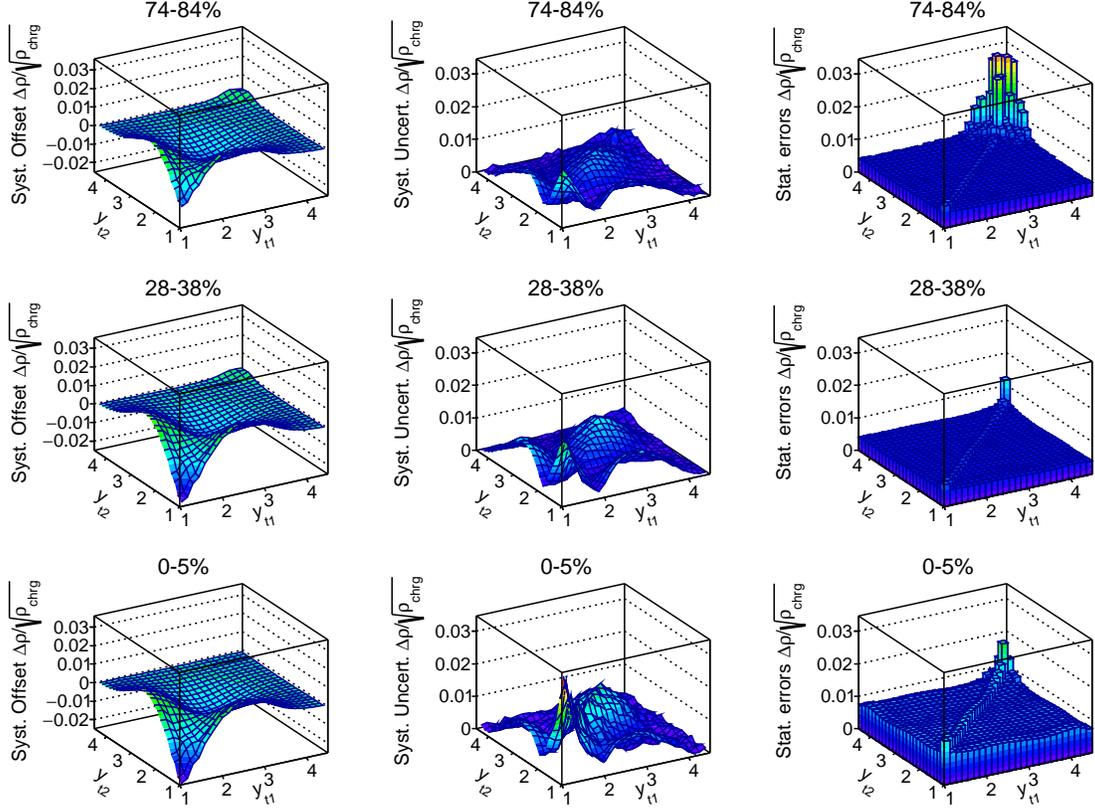}
\caption{\label{Fig8}
Systematic offsets, systematic uncertainties and statistical errors in the CI, all-azimuth correlations in columns of panels from left-to-right, respectively. Representative results for centralities 74-84\%, 28-38\% and 0-5\% are shown in rows of panels from upper to lower, respectively.}
\end{figure*}

\section{Comparison of statistical and systematic errors}
\label{Appendix-D}

	Statistical errors are compared with the corresponding systematic offsets and uncertainties in Fig.~\ref{Fig8} for centralities 74-84\%, 28-38\% and 0-5\%. Statistical errors are approximately constant over most of the binned $(k,l)$ space, increasing by about $\sqrt{2}$ along the main diagonal due to data symmetrization as discussed in Sec.~\ref{SecIII}. The statistical errors increase significantly towards higher values of $y_T$. The statistical errors for the 0-5\% and 5-9\% (not shown) centralities are larger than those errors for all other centrality bins because the two more-central bins include approximately one-half as many collision events as the other centralities. 

The systematic offsets do not change significantly with centrality while the systematic uncertainties generally increase. Systematic uncertainties are generally larger than statistical errors in the regions of interest where significant correlation structures appear, i.e. along the main diagonal and near the saddle-shape minima. Statistical errors dominate at higher $y_T \geq 4$ precluding investigation of possible correlation structures in this region with the present data set. 


\section{Resonance contributions}
\label{Appendix-B}

Correlations between the daughters of a short-lived, strongly decaying resonance in a high-energy, heavy-ion collision are generated by the decay dynamics itself and may contribute to the correlations presented in this paper. Resonance production, decay, and regeneration, as well as scattering of the resonance decay particles in the medium, are thought to occur in high-energy heavy-ion collisions~\cite{STARrho,PHENIXomega}. It is likely that some or all of the correlations between the daughter particles will be strongly dissipated in the medium. For the present estimate it was assumed that the number of decay pairs contributing to the final-state correlations on $(y_{T1},y_{T2})$ correspond to the surviving number of resonance decays estimated from the observed yields in the invariant mass distribution.

Resonance decay contributions to unidentified charged-particle correlations on transverse rapidity in the momentum range studied here are dominated by $\rho \rightarrow \pi^+ + \pi^-$ (BR $\approx$ 100\%) and $\omega \rightarrow \pi^+ + \pi^- + \pi^0$ (BR $\approx$ 89\%), based on analysis of the measured $\pi^+,\pi^-$ invariant mass distribution for peripheral Au+Au collisions~\cite{STARrho,PHENIXomega}. A Monte Carlo simulation was done where the measured $\rho$ and $\omega$ meson $p_T$ distributions~\cite{STARrho,PHENIXomega} were randomly sampled and the per-event yields were determined from measured $\rho^0/\pi^-$ and $\omega/\pi$ ratios~\cite{STARrho,PHENIXomega}. The CI, all-azimuth correlation quantity was calculated using the unidentified charged-particle pair reference distribution and prefactor as in Eq.~(\ref{Eq16}). The $\pi^+,\pi^-$ pairs from $\rho,\omega$ decays are primarily distributed within $y_T < 3$. The overall contributions to the correlation amplitudes varied from approximately 0.01 to 0.03 at low $y_T$ from peripheral to central collisions, respectively, and similarly from 0.007 to $-0.013$ in the region of the (3,3) correlation peak. The latter are about 8\% of the amplitude of the correlation peak in peripheral and more-central collisions. 

\end{document}